\documentstyle[prb,aps]{revtex}
\begin{document}
\draft

\author{Robert H. Gowdy\thanks{%
E-mail address: (Internet) rgowdy@cabell.vcu.edu}}
\address{Department of Physics,\\
Virginia Commonwealth University,\\
Richmond, Virginia 23284-2000}
\title{Affine projection tensor geometry: Lie derivatives and isometries}
\date{August 10, 1994;  gr-qc/9408014}
\maketitle

\begin{abstract}
The generalized projection-tensor geometry introduced in an earlier paper is
extended. A compact notation for families of projected objects is introduced
and used to summarize the results of the previous paper and obtain fully
projected decompositions of Lie derivatives of the projection tensor field,
the metric and the projected parts of the metric. These results are applied
to the analysis of spacetimes with isometries. The familiar cases of
spacetimes with isotropic group orbits --- cosmological models and spherical
symmetry --- are discussed as illustrations of the results.
\end{abstract}

\section{Introduction}

Many important problems in general relativity can be stated in terms of
projections onto subspaces and the relationship of those projections to
diffeomorphisms. From the earliest days of general relativity, exact
solutions to Einstein's equations have been sought by assuming isometry
groups --- metric preserving diffeomorphisms --- and projecting the metric
and other geometrical objects onto the orbits of those groups. In more
recent times, the emphasis has shifted to organizing Einstein's equations
for efficient numerical evolution of spacetimes from initial data. However,
the quantities which are needed for that endeavor are still the Lie
derivatives --- infinitesimal diffeomorphisms --- of tensors which have been
projected in various ways. Thus, the relationship between projections and
diffeomorphisms remains at the center of our attempts to understand the
dynamics of Einstein's field equations. This paper, the second in a series
on the geometry of projection-tensor fields, focuses on this relationship.
The previous paper, Ref. 1, worked out all of the ways in which projection
tensor fields can interact with a connection on a manifold. This paper
performs that same task for Lie derivatives.

The point of this series of papers is the flexibility of a projection tensor
geometry which is not restricted to normal projections or projections of
codimension one. Normal projections of codimension one are familiar from the
3+1 approach to the initial value problems of spacetime field theories.
Naturally, the conventional 3+1 results can be obtained by specializing the
results in this series but there are more direct ways to obtain those
results.\cite{walner} This series of papers attempts to place those results
in a wider context.

The definitions and results of projection tensor geometry are summarized in
Sec. \ref{prjctn.sum}. Although the content of Ref. 1 is summarized here,
this treatment is not intended to be self-contained and the reader who
wishes to understand this paper in detail should begin with Ref. 1. Section
\ref{assembly} introduces a notational convenience, {\em assemblies of
restricted tensors,} or {\em projection assemblies} which make projection
tensor expressions more compact, and identical in form to their counterparts
in unprojected Riemannian geometry. This compact new notation is used to
present the key results of Ref.~1. The main results of this paper are
presented in Sec. \ref{Lie} which evaluates the Lie derivatives of projected
geometrical structures and in Sec. \ref{orbits} which considers the
properties of projections onto the orbits of isometry groups.

As in Ref. 1, the paper closes with some familiar applications to provide a
context for these results. In particular, Sec. \ref{apps} revisits
Birkhoff's Theorem and shows how the results given here apply to spherically
symmetric systems. Some natural generalizations of spherical symmetry lead
to standard cosmological models as well as a five-dimensional Kaluza-Klein%
\cite{Kaluza-Klein-gen} cosmological model which displays a spontaneous
dimensional reduction. The interesting point about these examples is that
they are all fundamentally the same, differing only by the ''accidents'' of
dimension and metric signature.

\section{Summary of projection tensor geometry}
\label{prjctn.sum}

\subsection{Projection tensors and subspaces}

A {\em projection tensor field} $H$ assigns to each point $P$ of a manifold,
a map $H\left( P\right) :T_P\rightarrow T_P$ such that $H\left( P\right)
^2=H\left( P\right) $. The {\em complement} $V=I-H$ of a projection tensor
field is also a projection tensor field. The pull-backs $H\left( P\right)
^{*}:\hat T_P\rightarrow \hat T_P$ and $V\left( P\right) ^{*}:\hat T%
_P\rightarrow \hat T_P$ are used to project one-forms. The tensor products
of the subspaces $H\left( P\right) T_P,\ V\left( P\right) T_P$,$\ H\left(
P\right) ^{*}\hat T_P,\ V\left( P\right) ^{*}\hat T_P$ are called {\em fully
projected tensor subspaces}. Each fully projected subspace $S_P$ may be
characterized by a projection operator $O\left( P\right) $ which takes any
tensor $M$ of the appropriate rank into a tensor $O\left( P\right) M$ which
is in $S_P$ and which acts as an identity operator on $S_P$. For example, $H$
itself belongs to the fully projected subspace $H\left( P\right) ^{*}\hat T%
_P\otimes H\left( P\right) T_P$ and the corresponding projection operator $%
O\left[ ^H{}_H\right] $ takes any rank-$%
{1 \choose 1}
$ tensor $M$ into a tensor in that subspace:%
$$
\left( O\left[ ^H{}_H\right] M\right) {}^\alpha {}_\beta =H^\alpha {}_\rho
{}H^\sigma {}_\beta M^\rho {}_\sigma .
$$
A tensor field which has values only in fully projected tensor subspaces
will be referred to as a {\em restricted tensor field}. Projection tensor
geometry seeks to express everything in terms of such restricted tensor
fields.

The term ''restricted tensor field'' is new in this paper. It replaces the
term ''fully projected tensor field'' which was used in Ref. 1. This change
in terminology is needed because many restricted tensor fields have
definitions which are restricted to a given fully projected subspace and are
not simply projections of their counterparts in unprojected spacetime
geometry.

Projections of higher-rank tensor fields can be cumbersome in standard index
notation. I will use an abbreviated notation which places symbols for the
projection tensors in a pattern to indicate where they would act in an index
notation. The projection operator $O\left[ ^H{}_H\right] $ shown above is
one example. Another example is the projection
$$
\left( O\left[ _H{}^V{}_H\right] \,M\right) {}_\alpha {}^\beta {}_\gamma
=M\left[ _H{}^V{}_H\right] {}_\alpha {}^\beta {}_\gamma =H^\rho {}_\alpha
{}V^\beta {}_\sigma {}H^\tau {}_\gamma {}M_\rho {}^\sigma {}_\tau .
$$
of a tensor $M$ into the tensor $M\left[ _H{}^V{}_H\right] $ which belongs
to the fully projected subspace%
$$
T\left[ _H{}^V{}_H\right] _P=H\left( P\right) ^{*}\hat T_P\otimes V\left(
P\right) T_P\otimes H\left( P\right) ^{*}\hat T_P.
$$

\subsection{Projection curvatures}

For each projection tensor field $H$, there is a {\em projection curvature
tensor}
\begin{equation}
\label{prjctn.crv.def}h_H{}^\gamma {}_{\alpha \beta }{}=H^\rho {}_\alpha
\left( H^\delta {}_\beta \nabla _\delta H^\gamma {}_\rho \right) =\left(
\nabla H\left[ ^I{}_{H\ H}\right] \right) {}^\gamma {}_{\alpha \beta }
\end{equation}
and a {\em transpose projection curvature tensor}
\begin{equation}
\label{trans.prjctn.crv.def}h_H^T{}_\gamma {}^\alpha {}_\beta {}=H^\alpha
{}_\rho \left( H^\delta {}_\beta \nabla _\delta H^\rho {}_\gamma \right)
=\left( \nabla H\left[ ^H{}_{I\ H}\right] \right) {}^\gamma {}_{\alpha \beta
}.
\end{equation}
These tensor fields obey the projection identities
\begin{equation}
\label{prjctn.crv.ids}h_H\left[ ^V{}_{H\ H}{}\right] {}^\gamma {}_{\alpha
\beta }{}=h_H{}^\gamma {}_{\alpha \beta }{},\qquad h_H^T{}\left[
{}_V{}^H{}_{H\ }\right] {}_\gamma {}^\alpha {}_\beta {}=h_H^T{}_\gamma
{}^\alpha {}_\beta {}
\end{equation}
and are therefore restricted. Since the complement $V$ of $H$ is also a
projection tensor field, it too has projection curvatures which obey
identities which are the complements of the ones given above. The projection
curvatures may be used to expand the covariant derivative of the projection
tensor $H$ in terms of restricted tensor fields.
\begin{equation}
\label{prjctn.grad.exp}\nabla _\delta H^\alpha {}_\beta =h_H{}^\alpha
{}_{\beta \delta }{}-h_V{}^\alpha {}_{\beta \delta }{}+h_H^T{}_\beta
{}^\alpha {}_\delta {}-h_V^T{}_\beta {}^\alpha {}_\delta {}
\end{equation}
Some tensors derived from the projection curvatures are:%
$$
\begin{array}{ll}
\text{Divergence} & \theta _H^T{}_\beta \quad =\;h_H^T{}_\beta {}^\rho
{}_\rho {} \\
\text{Twist} & \omega _H{}^\alpha {}_{\beta \delta }{}=h_H{}^\alpha
{}_{\left[ \beta \delta \right] }{} \\
\text{Expansion} & \theta _H{}^\alpha {}_{\beta \delta }{}\,=h_H{}^\alpha
{}_{\left( \beta \delta \right) }{}
\end{array}
$$

\subsection{Restricted derivatives}

For a restricted tensor field $M,$ several {\em projected derivatives} can
be defined. If $O$ is the projection operator which characterizes the
subspace which $M$ belongs to, then the {\em projected covariant derivative}
of $M$ is defined to be the tensor with components%
$$
D_\delta \,M=O\,\nabla _\delta \,M.
$$
Here, the indexes which are not associated with the derivative have been
suppressed. Since $D_\delta \,M$ does not, itself, belong to a fully
projected tensor subspace, it is useful to define derivatives which do. The
{\em restricted covariant derivatives} of $M$ are defined to be%
$$
D_{H\,\delta }\,M=H^\sigma {}_\delta \,D_\sigma \,M,\qquad D_{V\,\delta
}\,M=V^\sigma {}_\delta \,D_\sigma \,M.
$$

The full decomposition of the covariant derivative of a fully projected
tensor can be expressed in terms of its restricted derivatives and the
projection curvatures. For a vector field $v$ with $v\left( P\right) \in
HT_P $, the covariant derivative has the decomposition
\begin{equation}
\label{covderiv.H.v}\nabla _{H\,\delta }\,v^\alpha =H^\sigma {}_\delta
\,\nabla _\sigma v^\alpha =D_{H\,\delta }v^\alpha +h_H{}^\alpha {}_{\rho
\delta }{}v^\rho
\end{equation}
\begin{equation}
\label{covderiv.V.v}\nabla _{V\,\delta }\,v^\alpha =V^\sigma {}_\delta
\,\nabla _\sigma v^\alpha =D_{V\,\delta }v^\alpha -h_V^T{}_\rho {}^\alpha
{}_\delta {}v^\rho
\end{equation}
The covariant derivative of a vector field $v$ with $v\left( P\right) \in
VT_P$ is given by the complement of the above decomposition. For a one-form
field $\varphi $ with $\varphi \left( P\right) \in H^{*}\hat T_P$ the
covariant derivative has the decomposition
\begin{equation}
\label{covderiv.H.phi}\nabla _{H\,\delta }\,\varphi _\alpha =D_{H\,\delta
}\,\varphi _\alpha +h_H^T{}_\alpha {}^\rho {}_\delta {}\varphi _\rho
\end{equation}
\begin{equation}
\label{covderiv.V.phi}\nabla _{V\,\delta }\,\varphi _\alpha =D_{V\,\delta
}\,\varphi _\alpha -h_V{}{}^\rho {}_{\alpha \delta }{}\varphi _\rho
\end{equation}
and the complementary equations give the decomposition of a one-form field
with the complementary projection property. In general, for an arbitrary
rank fully projected tensor, $M$, the covariant derivative decomposition
takes the form%
$$
\nabla _{X\,\delta }\,M=D_{X\,\delta }\,M+\text{corrections.}
$$
where $X$ can be either $H$ or $V$. There is a correction for each index on
the tensor. Each correction consists of one of the two tensors $h_X$ or $%
h_X^T$ with either its first or its last index contracted with an index on $%
M $. Because of the projection curvature identities (Eq. (\ref
{prjctn.crv.ids})) there will always be just one way to form such a term.
The sign of each term is positive when the tensor index being contracted is
projected in the same way as the differentiating index and negative
otherwise.

\subsection{Restricted metric and metricity}

The form-metric $g^{\mu \nu }$ is decomposed into the restricted tensors $%
g^{XY}{}^{\mu \nu }{}=g\left[ ^{XY}\right] {}^{\mu \nu }$ where the
projection labels $X,Y$ stand for either $H$ or $V$. Similarly, the
metricity tensor $Q^{\mu \nu }{}_\rho =-\nabla _\rho g^{\mu \nu }$ may be
decomposed into restricted parts according to the following expressions
\begin{equation}
\label{intr.metricity}Q\left[ ^{HH}{}_H\right] {}^{\mu \nu }{}_\delta
{}=Q_H^{HH}{}^{\mu \nu }{}_\delta {}+g^{HV}{}^{\mu \rho }{}h_H^T{}_\rho
{}^\nu {}_\delta {}+g^{VH}{}^{\rho \nu }{}h_H^T{}_\rho {}^\mu {}_\delta {}
\end{equation}
\begin{equation}
\label{fermi.metricity}Q\left[ ^{HH}{}_V\right] {}^{\mu \nu }{}_\delta
{}=Q_V^{HH}{}^{\mu \nu }{}_\delta {}-g^{HV}{}^{\mu \rho }{}h_V{}^\nu
{}_{\rho \delta }{}-g^{VH}{}^{\rho \nu }{}h_V{}^\mu {}_{\rho \delta }{}
\end{equation}
\begin{equation}
\label{hT-h.relation}Q\left[ ^{HV}{}_H\right] {}^{\mu \nu }{}_\delta
{}=Q_H^{HV}{}^{\mu \nu }{}_\delta {}-g^{HH}{}^{\mu \rho }{}h_H{}^\nu
{}_{\rho \delta }{}+g^{VV}{}^{\rho \nu }{}h_H^T{}_\rho {}^\mu {}_\delta {}
\end{equation}
\begin{equation}
\label{hTV-h.relation}Q\left[ ^{HV}{}_V\right] {}^{\mu \nu }{}_\delta
{}=Q_V^{HV}{}^{\mu \nu }{}_\delta {}+g^{HH}{}^{\mu \rho }{}h_V^T{}_\rho
{}^\nu {}_\delta {}-g^{VV}{}^{\rho \nu }{}h_V{}^\mu {}_{\rho \delta }{}
\end{equation}
and their complements. Here the {\em restricted metricities}
$$
Q_Z^{XY}{}^{\mu \nu }{}_\delta {}=-D_{Z\,\delta }\,g^{XY}{}^{\mu \nu }
$$
include the {\em intrinsic metricity} $Q_H^{HH}$ associated with the
subspaces $HT_P$, the intrinsic metricity $Q_V^{VV}$ associated with the
subspaces $VT_P$ and a collection of unfamiliar objects such as $Q_H^{HV}$
which I choose to call the {\em cross-projected metricities}.

Notice that the restricted metric tensors $g^{XY}{}^{\mu \nu }$ are just the
projections $g\left[ ^{XY}\right] {}^{\mu \nu }$ of the spacetime tensor
field $g^{\mu \nu }$ but the restricted metricities $Q_Z^{XY}{}^{\mu \nu
}{}_\delta $ are defined by using restricted covariant derivatives within
each fully projected tensor subspace and are not just projections of the
full spacetime metricity.

\subsection{Restricted torsion}

The torsion tensor $S^\rho {}_{\mu \nu }{}$ is defined through the
commutator of covariant derivatives acting on a function $f$ according to
\begin{equation}
\label{torsion.def}\left[ \nabla _\nu ,\nabla _\mu \right] f=S^\rho {}_{\mu
\nu }\nabla _\rho f.
\end{equation}
It can be decomposed into restricted tensor fields by defining the {\em %
restricted torsion tensors} $S_{XY}^Z{}^\rho {}_{\mu \nu }$ according to
\begin{equation}
\label{restrict.torison.def}\left[ D_{X\nu },D_{Y\mu }\right]
f=S_{XY}^H{}^\rho {}_{\mu \nu }D_{H\rho }f+S_{XY}^V{}^\rho {}_{\mu \nu
}D_{V\rho }f
\end{equation}
with the resulting decomposition
\begin{equation}
\label{int.tor.def}S\left[ ^H{}_{H~H}\right] {}^\rho {}_{\mu \nu
}{}=S_{HH}^H{}^\rho {}_{\mu \nu }{}
\end{equation}
\begin{equation}
\label{vhh.tor.def}S\left[ ^V{}_{H~H}\right] {}^\rho {}_{\mu \nu
}{}=S_{HH}^V{}^\rho {}_{\mu \nu }{}-2h_H{}^\rho {}_{\left[ \mu \nu \right]
}{}
\end{equation}
\begin{equation}
\label{hhv.tor.def}S\left[ ^H{}_{H~V}\right] {}^\rho {}_{\mu \nu
}{}=S_{HV}^H{}^\rho {}_{\mu \nu }{}-h_H^T{}_\nu {}^\rho {}_\mu {}
\end{equation}
as well as the complements of these expressions.

Notice once again, the critical distinction between restricted objects and
the projections of the corresponding spacetime objects. In this case, the
restricted torsions are clearly different from the projections of the full
spacetime torsion.

\section{Projection assembly notation}
\label{assembly}

\subsection{Organizing collections of restricted tensors}

\subsubsection{Tensor index - projection label pairs}

A restricted tensor such as the restricted torsion described above carries a
{\em projection label} for each of its tensor indexes. Thus, $%
S_{XY}^Z{}^\rho {}_{\mu \nu }$ carries the label $Z$ for the tensor index $%
\rho $, the label $X$ for the tensor index $\mu $ and the label $Y$ for the
tensor index $\nu $. The projection labels stand for projection tensors and
indicate the projection identities which are associated with each tensor
index. For example, the restricted torsion obeys the identity%
$$
X^\sigma {}_\mu S_{XY}^Z{}^\rho {}_{\sigma \nu }=S_{XY}^Z{}^\rho {}_{\mu \nu
}
$$
as well as two more identities associated with its other indexes.

Represent each index-label pair by a single compound index so that the
restricted torsion tensors become%
$$
S^{\left\langle Z\rho \right\rangle }{}_{\left\langle X\mu \right\rangle
\left\langle Y\nu \right\rangle }=S_{XY}^Z{}^\rho {}_{\mu \nu }.
$$
Abbreviate even more and use a single symbol $\left\langle \rho
\right\rangle $ to stand for an index-label pair $\left\langle X\rho
\right\rangle $ and interpret the summation convention on a repeated
abbreviated symbol to imply a sum over both the visible index value $\rho $
and the invisible projection-label $X$. In this compact notation, the
definition (Eq. \ref{restrict.torison.def}) of the restricted torsion
tensors becomes just%
$$
\left[ D_{\left\langle \nu \right\rangle },D_{\left\langle \mu \right\rangle
}\right] f=S{}^{\left\langle \rho \right\rangle }{}_{\left\langle \mu
\right\rangle \left\langle \nu \right\rangle }D_{\left\langle \rho
\right\rangle }f.
$$

When an unrestricted tensor such as the full torsion tensor is projected,
the result is also an object with corresponding projection labels and
indexes and it too can be represented in terms of index-label pair indexes.
Thus, the projections of the full torsion tensor can be represented as%
$$
S\left[ \;\right] {}^{\left\langle Z\rho \right\rangle }{}_{\left\langle
X\mu \right\rangle \left\langle Y\nu \right\rangle }=S\left[
^Z{}_{X~Y}\right] {}^\rho {}_{\mu \nu }
$$
where the empty square bracket remains in order to distinguish these
projections from the family of restricted torsion tensors.

\subsubsection{Assemblies of restricted tensors}

When a collection of restricted tensors is organized into a single object
with projection labels, the result is a larger geometrical object which I
choose to call an {\em assembly}. Thus, the index-label pair components $%
S{}^{\left\langle \rho \right\rangle }{}_{\left\langle \mu \right\rangle
\left\langle \nu \right\rangle }$ are regarded as specifying the {\em %
restricted torsion assembly} while the components $S\left[ \;\right]
{}^{\left\langle \rho \right\rangle }{}_{\left\langle \mu \right\rangle
\left\langle \nu \right\rangle }$ are regarded as specifying the {\em %
projected torsion assembly}. Similarly, the array of restricted metricity
components $Q{}^{\left\langle \mu \right\rangle \left\langle \nu
\right\rangle }{}_{\left\langle \delta \right\rangle }$ specify the {\em %
restricted metricity assembly} while the projections $Q\left[ \;\right]
{}^{\left\langle \mu \right\rangle \left\langle \nu \right\rangle
}{}_{\left\langle \delta \right\rangle }$ specify the {\em projected
metricity assembly}. The advantage of organizing restricted tensors into
assemblies is that the index-label pair notation can then be used to make
very compact expressions which are usually identical in structure to
familiar unprojected tensor expressions.

\subsubsection{Projection gradient assembly}

It is particularly convenient to organize the projection curvatures into a
single assembly, the {\em projection gradient assembly}. In terms of this
object, the projection gradient decomposition (Eq. \ref{prjctn.grad.exp})
becomes just
\begin{equation}
\label{grad.h(k)}\nabla H\left[ \;\right] {}^{\left\langle \beta
\right\rangle }{}_{\left\langle \alpha \right\rangle \left\langle \delta
\right\rangle }=\nabla H{}^{\left\langle \beta \right\rangle
}{}_{\left\langle \alpha \right\rangle \left\langle \delta \right\rangle }{}
\end{equation}
and the decomposition of the covariant derivative of a vector field $v\in
HT_P$ (compare to Eqs. \ref{covderiv.H.v} and \ref{covderiv.V.v}) is
\begin{equation}
\label{grad.Hv(k)}\nabla _{\left\langle \delta \right\rangle
}v^{\left\langle \alpha \right\rangle }=D_{\left\langle \delta \right\rangle
}v^{\left\langle \alpha \right\rangle }+\nabla H{}^{\left\langle \alpha
\right\rangle }{}_{\left\langle \rho \right\rangle \left\langle \delta
\right\rangle }{}v^{\left\langle \rho \right\rangle }
\end{equation}
while, for a form-field $\varphi \in H\hat T_P$ (compare to Eqs. \ref
{covderiv.H.phi} and \ref{covderiv.V.phi})
\begin{equation}
\label{grad.Hphi(k)}\nabla _{\left\langle \delta \right\rangle }\varphi
_{\left\langle \beta \right\rangle }=D_{\left\langle \delta \right\rangle
}\varphi _{\left\langle \beta \right\rangle }+\varphi _{\left\langle \rho
\right\rangle }\nabla H{}^{\left\langle \rho \right\rangle }{}_{\left\langle
\beta \right\rangle \left\langle \delta \right\rangle }{}.
\end{equation}

\subsubsection{The complementation tensor}

{}From Eq. \ref{grad.h(k)} the complement of the assembly $\nabla H$ is $%
\nabla V=-\nabla H$. so that the covariant derivative of a vector field $%
v\in VT_P$ is given by Eq. \ref{grad.Hv(k)} with the sign of $\nabla H$
reversed. A similar reversal occurs for form-fields. In order to express
this reversal in expressions for arbitrary vector or form fields, introduce
the {\em complementation tensor}%
$$
C^\sigma {}_\rho =H^\sigma {}_\rho -V^\sigma {}_\rho
$$
and the corresponding assembly so that , for any restricted vector field $v$
and form-field $\varphi $,
\begin{equation}
\label{grad.v(k)}\nabla _{\left\langle \delta \right\rangle }v^{\left\langle
\alpha \right\rangle }=D_{\left\langle \delta \right\rangle }v^{\left\langle
\alpha \right\rangle }+\nabla H{}^{\left\langle \alpha \right\rangle
}{}_{\left\langle \sigma \right\rangle \left\langle \delta \right\rangle
}{}C^{\left\langle \sigma \right\rangle }{}_{\left\langle \rho \right\rangle
}v^{\left\langle \rho \right\rangle }
\end{equation}
\begin{equation}
\label{grad.phi(k)}\nabla _{\left\langle \delta \right\rangle }\varphi
_{\left\langle \beta \right\rangle }{}=D_{\left\langle \delta \right\rangle
}\varphi _{\left\langle \beta \right\rangle }{}+\varphi _{\left\langle
\sigma \right\rangle }C^{\left\langle \sigma \right\rangle }{}_{\left\langle
\rho \right\rangle }{}\nabla H{}^{\left\langle \rho \right\rangle
}{}_{\left\langle \beta \right\rangle \left\langle \delta \right\rangle }{}.
\end{equation}
It is no longer necessary to specify which subspace the vector $v$ or the
form $\varphi $ is restricted to --- only that it is restricted. When $V$ is
a timelike projection, the complementation tensor is just a representation
of the time reversal operator. The covariant derivative of the
complementation tensor follows from Eq. \ref{grad.h(k)}:%
$$
\nabla C\left[ \;\right] {}^{\left\langle \beta \right\rangle
}{}_{\left\langle \alpha \right\rangle \left\langle \delta \right\rangle
}=2\nabla H{}^{\left\langle \beta \right\rangle }{}_{\left\langle \alpha
\right\rangle \left\langle \delta \right\rangle }{}.
$$

\subsubsection{The projection curvature assembly}

The complementation and projection gradient assemblies obey the identity
\begin{equation}
\label{KC.identity}\nabla H^{\left\langle \alpha \right\rangle
}{}_{\left\langle \tau \right\rangle \left\langle \delta \right\rangle
}C^{\left\langle \tau \right\rangle }{}_{\left\langle \sigma \right\rangle
}{}=-C^{\left\langle \alpha \right\rangle }{}_{\left\langle \rho
\right\rangle }{}\nabla H{}^{\left\langle \rho \right\rangle
}{}_{\left\langle \sigma \right\rangle \left\langle \delta \right\rangle }{}
\end{equation}
which suggests that Eqs. \ref{grad.v(k)} and \ref{grad.phi(k)} can be
simplified and made to appear more like the normal relationships between
different types of derivative operators by defining the assembly
\begin{equation}
\label{CC.identity}K{}^{\left\langle \alpha \right\rangle }{}_{\left\langle
\sigma \right\rangle \left\langle \delta \right\rangle }=\nabla
H^{\left\langle \alpha \right\rangle }{}_{\left\langle \tau \right\rangle
\left\langle \delta \right\rangle }C^{\left\langle \tau \right\rangle
}{}_{\left\langle \sigma \right\rangle }{}=-C^{\left\langle \alpha
\right\rangle }{}_{\left\langle \rho \right\rangle }{}\nabla
H{}^{\left\langle \rho \right\rangle }{}_{\left\langle \sigma \right\rangle
\left\langle \delta \right\rangle }{}.
\end{equation}
The components of this object are
\begin{equation}
\label{k.def}
\begin{array}{ll}
K{}^{\left\langle H\beta \right\rangle }{}_{\left\langle H\alpha
\right\rangle \left\langle H\delta \right\rangle }{}=0,\qquad &
K{}^{\left\langle H\beta \right\rangle }{}_{\left\langle V\alpha
\right\rangle \left\langle H\delta \right\rangle }{}=-h_H^T{}_\alpha
{}^\beta {}_\delta {} \\
K{}^{\left\langle V\beta \right\rangle }{}_{\left\langle H\alpha
\right\rangle \left\langle H\delta \right\rangle }{}=h_H{}^\beta {}_{\alpha
\delta }{},\qquad & K{}^{\left\langle V\beta \right\rangle }{}_{\left\langle
V\alpha \right\rangle \left\langle H\delta \right\rangle }{}=0 \\
K{}^{\left\langle H\beta \right\rangle }{}_{\left\langle H\alpha
\right\rangle \left\langle V\delta \right\rangle }{}=0,\qquad &
K{}^{\left\langle H\beta \right\rangle }{}_{\left\langle V\alpha
\right\rangle \left\langle V\delta \right\rangle }{}=h_V{}^\beta {}_{\alpha
\delta }{} \\
K{}^{\left\langle V\beta \right\rangle }{}_{\left\langle H\alpha
\right\rangle \left\langle V\delta \right\rangle }{}=-h_V^T{}_\alpha
{}^\beta {}_\delta {}, & K{}^{\left\langle V\beta \right\rangle
}{}_{\left\langle V\alpha \right\rangle \left\langle V\delta \right\rangle
}{}=0
\end{array}
\end{equation}
This assembly does not change sign under complementation and plays a
satisfyingly familiar role as the generator of correction terms in the
relationship between covariant and restricted derivatives. For a restricted
tensor $m^{\alpha _1\alpha _2\ldots \alpha _n}{}_{\beta _1\beta _2\ldots
\beta _n}$ of arbitrary rank, the decomposition of the gradient into
restricted parts is just
\begin{equation}
\label{del.tensor.prjct}
\begin{array}{c}
\left( \nabla m\right) \left[ \,\right] ^{\left\langle \alpha
_1\right\rangle \ldots \left\langle \alpha _n\right\rangle }{}_{\left\langle
\beta _1\right\rangle \ldots \left\langle \beta _n\right\rangle
;\left\langle \delta \right\rangle }=D_{\left\langle \delta \right\rangle
}m^{\left\langle \alpha _1\right\rangle \ldots \left\langle \alpha
_n\right\rangle }{}_{\left\langle \beta _1\right\rangle \ldots \left\langle
\beta _n\right\rangle } \\
+m^{\left\langle \sigma \right\rangle \ldots \left\langle \alpha
_n\right\rangle }{}_{\left\langle \beta _1\right\rangle \ldots \left\langle
\beta _n\right\rangle }K{}^{\left\langle \alpha _1\right\rangle
}{}_{\left\langle \sigma \right\rangle \left\langle \delta \right\rangle
}+\ldots +m^{\left\langle \alpha _1\right\rangle \ldots \left\langle \sigma
\right\rangle }{}_{\left\langle \beta _1\right\rangle \ldots \left\langle
\beta _n\right\rangle }K{}^{\left\langle \alpha _n\right\rangle
}{}_{\left\langle \sigma \right\rangle \left\langle \delta \right\rangle }
\\
-m^{\left\langle \alpha _1\right\rangle \ldots \left\langle \alpha
_n\right\rangle }{}_{\left\langle \sigma \right\rangle \ldots \left\langle
\beta _n\right\rangle }K{}^{\left\langle \sigma \right\rangle
}{}_{\left\langle \beta _1\right\rangle \left\langle \delta \right\rangle
}{}-\ldots -m^{\left\langle \alpha _1\right\rangle \ldots \left\langle
\alpha _n\right\rangle }{}_{\left\langle \beta _1\right\rangle \ldots
\left\langle \sigma \right\rangle }K{}^{\left\langle \sigma \right\rangle
}{}_{\left\langle \beta _n\right\rangle \left\langle \delta \right\rangle
}{}.
\end{array}
\end{equation}

The metricity tensor assembly can now be calculated directly with a result,
\begin{equation}
\label{metricity(k)}Q\left[ \;\right] {}^{\left\langle \mu \right\rangle
\left\langle \nu \right\rangle }{}_{\left\langle \delta \right\rangle
}{}=Q{}^{\left\langle \mu \right\rangle \left\langle \nu \right\rangle
}{}_{\left\langle \delta \right\rangle }{}-2g{}^{\left\langle \rho
\right\rangle (\left\langle \mu \right\rangle }K{}^{\left\langle \nu
\right\rangle )}{}_{\left\langle \rho \right\rangle \left\langle \delta
\right\rangle }{}
\end{equation}
which is equivalent to Eqs. \ref{intr.metricity},\ref{fermi.metricity},\ref
{hT-h.relation}, and \ref{hTV-h.relation} as well as their complements.
Similarly, the torsion assembly becomes
\begin{equation}
\label{torsion(k)}S\left[ \;\right] {}^{\left\langle \rho \right\rangle
}{}_{\left\langle \mu \right\rangle \left\langle \nu \right\rangle
}{}=S{}^{\left\langle \rho \right\rangle }{}_{\left\langle \mu \right\rangle
\left\langle \nu \right\rangle }{}-2K{}^{\left\langle \rho \right\rangle
}{}_{\left[ \left\langle \mu \right\rangle \left\langle \nu \right\rangle
\right] }{}
\end{equation}
which is equivalent to Eqs. \ref{int.tor.def}, \ref{vhh.tor.def} and \ref
{hhv.tor.def}. as well as their complements.

\subsubsection{Assemblies as representations of tensors}

The reason for emphasizing the distinction between restricted and projected
objects should now be apparent. When a collection of restricted tensor
fields is actually the set of all projections of a spacetime tensor, the
corresponding assembly is just a representation of the {\em original
spacetime tensor field}. For example, the metric assembly $g^{\left\langle
\mu \right\rangle \left\langle \nu \right\rangle }$ is simply a
representation of the spacetime metric tensor $g^{\mu \nu }$. In an adapted
frame, one would refer to this representation as ''partitioning the matrix
of coefficients''. Similarly, the projected torsion assembly $S\left[
\;\right] {}^{\left\langle \rho \right\rangle }{}_{\left\langle \mu
\right\rangle \left\langle \nu \right\rangle }$ is just a representation of
the spacetime torsion tensor $S{}^\rho {}_{\mu \nu }$ but the restricted
torsion assembly $S{}^{\left\langle \rho \right\rangle }{}_{\left\langle \mu
\right\rangle \left\langle \nu \right\rangle }$ is not.

The concept of an assembly as a representation of a spacetime tensor can be
made explicit by introducing a set of basis vectors $e_\mu $ on the full
tangent space $T_P$ and noticing that the vectors
$$
e_{\left\langle H\mu \right\rangle }=He_\mu ,\qquad e_{\left\langle V\mu
\right\rangle }=Ve_\mu
$$
form an {\em overcomplete basis} for the subspaces $HT_P$ and $VT_P$
respectively. Similarly, the forms%
$$
\omega ^{\left\langle H\nu \right\rangle }=H^{*}\omega ^\nu ,\qquad \omega
^{\left\langle V\nu \right\rangle }=V^{*}\omega ^\nu
$$
are overcomplete bases for the subspaces $H^{*}\hat T_P$ and $V^{*}\hat T_P$%
. An assembly can be thought of as the expansion of a tensor in terms of
these overcomplete sets of basis vectors. For example, the vector assembly $%
v^{\left\langle \alpha \right\rangle }$ which has components $%
v^{\left\langle H\alpha \right\rangle }$ and $v^{\left\langle V\alpha
\right\rangle }$ in $HT_P$ and $VT_P$ respectively may also be regarded as a
representation of the vector%
$$
v=v^{\left\langle \rho \right\rangle }e_{\left\langle \rho \right\rangle
}=v^{\left\langle H\rho \right\rangle }e_{\left\langle H\rho \right\rangle
}+v^{\left\langle V\rho \right\rangle }e_{\left\langle V\rho \right\rangle
}=v^H+v^V
$$
in the full tangent space $T_P$.

\subsection{Restricted curvatures}

\subsubsection{Definition in terms of assemblies}

The {\em restricted curvature tensors} are defined by the action of
restricted covariant derivatives on restricted vector fields according to%
$$
\left( \left[ D_{\left\langle \beta \right\rangle },D_{\left\langle \alpha
\right\rangle }\right] -S{}^{\left\langle \rho \right\rangle
}{}_{\left\langle \alpha \right\rangle \left\langle \beta \right\rangle
}D_{\left\langle \rho \right\rangle }\right) v^{\left\langle \gamma
\right\rangle }=v^{\left\langle \rho \right\rangle }R{}_{\left\langle \rho
\right\rangle }{}^{\left\langle \gamma \right\rangle }{}_{\left\langle
\alpha \right\rangle \left\langle \beta \right\rangle }.
$$
Because the restricted derivatives always project back into the same
projected subspace, this definition implies that the first two index-label
pairs of the restricted curvature must have the same label. In the notation
of Ref. 1, the restricted curvature tensors are%
$$
R{}_{\left\langle Z\rho \right\rangle }{}^{\left\langle Z\gamma
\right\rangle }{}_{\left\langle X\alpha \right\rangle \left\langle Y\beta
\right\rangle }=R_{XY}^Z{}_\rho {}^\gamma {}_{\alpha \beta }.
$$
For example, if the projection tensor $H$ projects vectors into the space
tangent to a spacelike hypersurface, then the restricted curvature $%
R_{HH}^H{}_\rho {}^\gamma {}_{\alpha \beta }$ is the familiar Riemannian
curvature of the geometry intrinsic to the hypersurface. Similarly, when $V$
is surface-forming, $R_{VV}^V{}_\rho {}^\gamma {}_{\alpha \beta }$ is the
Riemannian curvature intrinsic to the surfaces which it forms. The remaining
restricted curvature tensors are unfamiliar objects which I call {\em %
cross-projection curvatures}.

In order to take full advantage of the similarity of the restricted
curvature definition to the full curvature definition, it is useful to let
the projection labels on the first two indexes take on all values by
stipulating that%
$$
R{}_{\left\langle H\rho \right\rangle }{}^{\left\langle V\gamma
\right\rangle }{}_{\left\langle \alpha \right\rangle \left\langle \beta
\right\rangle }=R{}_{\left\langle V\rho \right\rangle }{}^{\left\langle
H\gamma \right\rangle }{}_{\left\langle \alpha \right\rangle \left\langle
\beta \right\rangle }=0.
$$
This stipulation completes the definition of an assembly which I choose to
call the {\em restricted curvature assembly}.

\subsubsection{Identities of the restricted curvature assembly}

Now convert the results of Ref. 1 into this compact notation. The restricted
curvature assembly is found to obey the {\em assembled torsion Bianchi
identities}.
\begin{equation}
\label{T-Bianchi-asm}R{}_{[\left\langle \rho \right\rangle }{}^{\left\langle
\gamma \right\rangle }{}_{\left\langle \alpha \right\rangle \left\langle
\beta \right\rangle ]}+D_{[\left\langle \rho \right\rangle }S^{\left\langle
\gamma \right\rangle }{}_{\left\langle \alpha \right\rangle \left\langle
\beta \right\rangle ]}+S^{\left\langle \gamma \right\rangle
}{}_{\left\langle \sigma \right\rangle [\left\langle \rho \right\rangle
}S^{\left\langle \sigma \right\rangle }{}_{\left\langle \alpha \right\rangle
\left\langle \beta \right\rangle ]}=0
\end{equation}
and the {\em assembled curvature Bianchi identities}%
$$
D_{[\left\langle \rho \right\rangle }R_{\left| \left\langle \delta
\right\rangle \right| }{}^{\left\langle \gamma \right\rangle
}{}_{\left\langle \alpha \right\rangle \left\langle \beta \right\rangle
]}{}+R_{\left\langle \delta \right\rangle }{}^{\left\langle \gamma
\right\rangle }{}_{\left\langle \sigma \right\rangle [\left\langle \rho
\right\rangle }{}S^{\left\langle \sigma \right\rangle }{}_{\left\langle
\alpha \right\rangle \left\langle \beta \right\rangle ]}{}=0.
$$
Each of these identities expands out to a family of relationships connecting
different restricted curvature tensors. The assembled torsion Bianchi
identities in particular are useful for expressing the unfamiliar
cross-projected curvature tensors in terms of more familiar objects.

In this same compact notation, the curvature-metricity identity which was
discussed in Ref. 1 becomes
\begin{equation}
\label{cmet.id}2g^{\left\langle \rho \right\rangle (\left\langle \delta
\right\rangle }R{}_{\left\langle \rho \right\rangle }{}^{\left\langle \gamma
\right\rangle )}{}_{\left\langle \alpha \right\rangle \left\langle \beta
\right\rangle }{}=2D_{[\left\langle \alpha \right\rangle
}{}Q{}^{\left\langle \gamma \right\rangle \left\langle \delta \right\rangle
}{}_{\left\langle \beta \right\rangle ]}{}+S{}^{\left\langle \rho
\right\rangle }{}_{\left\langle \alpha \right\rangle \left\langle \beta
\right\rangle }{}Q{}^{\left\langle \gamma \right\rangle \left\langle \delta
\right\rangle }{}_{\left\langle \rho \right\rangle }{}
\end{equation}
Notice that the intrinsic curvature assembly, the restricted torsion
assembly, and the restricted metricity assembly obey identities which have
exactly the same structure as the usual identities of the unprojected
curvature tensor.

\subsubsection{Generalized Gauss-Codazzi curvature projections}

A spectacular example of how the use of assemblies condenses complex
expressions is the projection decomposition of the curvature tensor. A
straightforward calculation from the definition of the full and restricted
Riemann tensors and the covariant derivative decompositions implied by Eqs. (%
\ref{grad.v(k)}) and (\ref{grad.phi(k)}) as well as an application of the
identities given in Eq.(\ref{KC.identity}) and (\ref{CC.identity}) yields
the expression%
$$
\begin{array}{c}
R\left[ \,\right] {}_{\left\langle \rho \right\rangle }{}^{\left\langle
\gamma \right\rangle }{}_{\left\langle \alpha \right\rangle \left\langle
\beta \right\rangle }=R{}_{\left\langle \rho \right\rangle }{}^{\left\langle
\gamma \right\rangle }{}_{\left\langle \alpha \right\rangle \left\langle
\beta \right\rangle }+2D_{[\left\langle \beta \right\rangle
}K{}^{\left\langle \gamma \right\rangle }{}_{\left| \left\langle \rho
\right\rangle \right| \left\langle \alpha \right\rangle
]}{}-S{}^{\left\langle \tau \right\rangle }{}_{\left\langle \alpha
\right\rangle \left\langle \beta \right\rangle }{}K{}^{\left\langle \gamma
\right\rangle }{}_{\left\langle \rho \right\rangle \left\langle \tau
\right\rangle }{} \\
-2K{}^{\left\langle \gamma \right\rangle }{}_{\left\langle \varsigma
\right\rangle [\left\langle \alpha \right\rangle }{}K{}^{\left\langle
\varsigma \right\rangle }{}_{\left| \left\langle \rho \right\rangle \right|
\left\langle \beta \right\rangle ]}{}.
\end{array}
$$
This one expression gives all of the projections of the curvature tensor in
terms of components of the intrinsic and extrinsic curvature assemblies. A
partial expansion of the full expression is useful because one finds two
basic sub-expressions:%
$$
R\left[ \;\right] {}_{\left\langle H\rho \right\rangle }{}^{\left\langle
H\gamma \right\rangle }{}_{\left\langle \alpha \right\rangle \left\langle
\beta \right\rangle }=R_{\left\langle H\rho \right\rangle }{}^{\left\langle
H\gamma \right\rangle }{}_{\left\langle \alpha \right\rangle \left\langle
\beta \right\rangle }-2K{}^{\left\langle \sigma \right\rangle
}{}_{\left\langle H\rho \right\rangle [\left\langle \alpha \right\rangle
}{}K{}^{\left\langle H\gamma \right\rangle }{}_{\left| \left\langle \sigma
\right\rangle \right| \left\langle \beta \right\rangle ]}{}
$$
$$
R\left[ \;\right] {}_{\left\langle H\rho \right\rangle }{}^{\left\langle
V\gamma \right\rangle }{}_{\left\langle \alpha \right\rangle \left\langle
\beta \right\rangle }=2D_{[\left\langle \beta \right\rangle
}K{}^{\left\langle V\gamma \right\rangle }{}_{\left| \left\langle H\rho
\right\rangle \right| \left\langle \alpha \right\rangle ]}{}-S^{\left\langle
\sigma \right\rangle }{}_{\left\langle \alpha \right\rangle \left\langle
\beta \right\rangle }K{}^{\left\langle V\gamma \right\rangle
}{}_{\left\langle H\rho \right\rangle \left\langle \sigma \right\rangle }{}
$$
which clearly generalize the Gauss-Codazzi equations for surface embedding.

\section{Lie derivatives}
\label{Lie}

\subsection{Basics}

\subsubsection{Standard Lie derivatives}

The Lie derivative\cite{Lie-Deriv-ref} of a vector field $v$ with respect to
a vector-field $N$, may be obtained from the commutator%
$$
\left( \pounds _Nv\right) f=\left[ N,v\right] f
$$
for any function, $f$. To relate the Lie derivative to the covariant
derivative, write the commutator in the form%
$$
\left( \pounds _Nv\right) ^\delta \nabla _\delta f=\left[ \left( N^\sigma
\nabla _\sigma \right) \left( v^\rho \nabla _\rho \right) -\left( v^\rho
\nabla _\rho \right) \left( N^\sigma \nabla _\sigma \right) \right] f.
$$
The basic relationship between the Lie derivative and the covariant
derivative then follows from the definition, Eq. \ref{torsion.def}, of the
torsion tensor. In terms of the quantity%
$$
\nabla _\rho ^{\prime }N^\delta =\nabla _\rho N^\delta -S^\delta {}_{\rho
\sigma }N^\sigma
$$
the relationship is found to be
$$
\left( \pounds _Nv\right) ^\delta =N^\sigma \nabla _\sigma v^\delta -v^\rho
\nabla _\rho ^{\prime }N^\delta .
$$
The Lie derivative of an arbitrary rank tensors $M$ is related to its
covariant derivatives by
\begin{equation}
\label{Lie.tensor.cov}
\begin{array}{c}
\left( \pounds _NM\right) ^{\delta _1\ldots \delta _n}{}_{\alpha _1\ldots
\alpha _n}=N^\sigma \nabla _\sigma M^{\delta _1\ldots \delta _n}{}_{\alpha
_1\ldots \alpha _n} \\
-M^{\rho \ldots \delta _n}{}_{\alpha _1\ldots \alpha _n}\nabla _\rho
^{\prime }N^{\delta _1}-\ldots -M^{\delta _1\ldots \rho }{}_{\alpha _1\ldots
\alpha _n}\nabla _\rho ^{\prime }N^{\delta _n} \\
+M^{\delta _1\ldots \delta _n}{}_{\rho \ldots \alpha _n}\nabla _{\alpha
_1}^{\prime }N^\rho +\ldots +M^{\delta _1\ldots \delta _n}{}_{\alpha
_1\ldots \rho }\nabla _{\alpha _n}^{\prime }N^\rho .
\end{array}
\end{equation}

\subsubsection{Restricted Lie derivatives}

A restricted Lie derivative can be defined in the same way that the
restricted covariant derivative was defined. For a restricted tensor-field $%
M $ which is characterized by the projection
\begin{equation}
\label{OM.identity}OM=M
\end{equation}
the restricted Lie derivative of $M$ is defined to be%
$$
L_NM=O\pounds _NM.
$$
Just as the restricted covariant derivatives of the projection tensor field $%
H$ are automatically zero, it is easy to see that%
$$
L_NH=0.
$$

The restricted Lie derivative of an assembly of restricted tensor fields $M$
is related to its restricted covariant derivatives by an expression with the
same structure as Eq. \ref{Lie.tensor.cov}
\begin{equation}
\label{rstrct.Lie.tensor.cov}
\begin{array}{c}
\left( L_NM\right) ^{\left\langle \delta _1\right\rangle \ldots \left\langle
\delta _n\right\rangle }{}_{\left\langle \alpha _1\right\rangle \ldots
\left\langle \alpha _n\right\rangle }=N^{\left\langle \sigma \right\rangle
}D_{\left\langle \sigma \right\rangle }M^{\left\langle \delta
_1\right\rangle \ldots \left\langle \delta _n\right\rangle }{}_{\left\langle
\alpha _1\right\rangle \ldots \left\langle \alpha _n\right\rangle } \\
-M^{\left\langle \rho \right\rangle \ldots \left\langle \delta
_n\right\rangle }{}_{\left\langle \alpha _1\right\rangle \ldots \left\langle
\alpha _n\right\rangle }D_{\left\langle \rho \right\rangle }^{\prime
}N^{\left\langle \delta _1\right\rangle }-\ldots -M^{\left\langle \delta
_1\right\rangle \ldots \left\langle \rho \right\rangle }{}_{\left\langle
\alpha _1\right\rangle \ldots \left\langle \alpha _n\right\rangle
}D_{\left\langle \rho \right\rangle }^{\prime }N^{\left\langle \delta
_n\right\rangle } \\
+M^{\left\langle \delta _1\right\rangle \ldots \left\langle \delta
_n\right\rangle }{}_{\left\langle \rho \right\rangle \ldots \left\langle
\alpha _n\right\rangle }D_{\left\langle \alpha _1\right\rangle }^{\prime
}N^{\left\langle \rho \right\rangle }+\ldots +M^{\left\langle \delta
_1\right\rangle \ldots \left\langle \delta _n\right\rangle }{}_{\left\langle
\alpha _1\right\rangle \ldots \left\langle \rho \right\rangle
}D_{\left\langle \alpha _n\right\rangle }^{\prime }N^{\left\langle \rho
\right\rangle }.
\end{array}
\end{equation}
where%
$$
D_{\left\langle \rho \right\rangle }^{\prime }N^{\left\langle \delta
\right\rangle }=D_{\left\langle \rho \right\rangle }N^{\left\langle \delta
\right\rangle }-S^{\left\langle \delta \right\rangle }{}_{\left\langle \rho
\right\rangle \left\langle \sigma \right\rangle }N^{\left\langle \sigma
\right\rangle }.
$$

\subsection{Restricted Lie derivatives of geometrical structures}

\subsubsection{The projection tensor}

Now work out the projection-tensor decomposition of \pounds $_NH$. Start
with the assembly-notation version of the spacetime expression.%
$$
\pounds _NH^{\left\langle \alpha \right\rangle }{}_{\left\langle \beta
\right\rangle }=N^{\left\langle \delta \right\rangle }\nabla _{\left\langle
\delta \right\rangle }H^{\left\langle \alpha \right\rangle }{}_{\left\langle
\beta \right\rangle }-H^{\left\langle \rho \right\rangle }{}_{\left\langle
\beta \right\rangle }\nabla _{\left\langle \rho \right\rangle }^{\prime
}N^{\left\langle \alpha \right\rangle }+H^{\left\langle \alpha \right\rangle
}{}_{\left\langle \rho \right\rangle }\nabla _{\left\langle \beta
\right\rangle }^{\prime }N^{\left\langle \rho \right\rangle }
$$
Use Eq. \ref{del.tensor.prjct} to represent the covariant derivatives in
terms of assemblies of restricted objects and use Eq. \ref{torsion(k)} to
eliminate the spacetime torsion tensor.

\begin{equation}
\label{Lie.H.asm}
\begin{array}{c}
\pounds _NH^{\left\langle \alpha \right\rangle }{}_{\left\langle \beta
\right\rangle }=H^{\left\langle \alpha \right\rangle }{}_{\left\langle \rho
\right\rangle }{}D_{\left\langle \beta \right\rangle }N^{\left\langle \rho
\right\rangle }-H^{\left\langle \rho \right\rangle }{}_{\left\langle \beta
\right\rangle }{}D_{\left\langle \rho \right\rangle }N^{\left\langle \alpha
\right\rangle } \\
+N^{\left\langle \sigma \right\rangle }\left( H^{\left\langle \rho
\right\rangle }{}_{\left\langle \beta \right\rangle }{}S{}^{\left\langle
\alpha \right\rangle }{}_{\left\langle \rho \right\rangle \left\langle
\sigma \right\rangle }{}-H^{\left\langle \alpha \right\rangle
}{}_{\left\langle \rho \right\rangle }{}S{}^{\left\langle \rho \right\rangle
}{}_{\left\langle \beta \right\rangle \left\langle \sigma \right\rangle
}{}\right)
\end{array}
\end{equation}
This expression is actually antisymmetric under complementation, as the
identity%
$$
\pounds _N\left( H+V\right) =\pounds _NH+\pounds _NV=0
$$
requires. It is useful to make this complementation antisymmetry manifest by
writing the expression in the form
\begin{equation}
\label{Lie.HC.asm}
\begin{array}{c}
\pounds _NH^{\left\langle \alpha \right\rangle }{}_{\left\langle \sigma
\right\rangle }=
\frac 12\left( C^{\left\langle \alpha \right\rangle }{}_{\left\langle \tau
\right\rangle }{}D_{\left\langle \sigma \right\rangle }N^{\left\langle \tau
\right\rangle }-C^{\left\langle \tau \right\rangle }{}_{\left\langle \sigma
\right\rangle }{}D_{\left\langle \tau \right\rangle }N^{\left\langle \alpha
\right\rangle }\right) \\ +\frac 12N^{\left\langle \varsigma \right\rangle
}\left( C^{\left\langle \tau \right\rangle }{}_{\left\langle \sigma
\right\rangle }{}S{}^{\left\langle \alpha \right\rangle }{}_{\left\langle
\tau \right\rangle \left\langle \varsigma \right\rangle }{}-C^{\left\langle
\alpha \right\rangle }{}_{\left\langle \tau \right\rangle
}{}S{}^{\left\langle \tau \right\rangle }{}_{\left\langle \sigma
\right\rangle \left\langle \varsigma \right\rangle }{}\right) .
\end{array}
\end{equation}

Simplify the result another way by finding its non-zero projections. Because
the projections of the full spacetime torsion are usually set to zero,
collect those terms together and write the result in the form
\begin{equation}
\label{Lie.H.HV.comp}\pounds _NH\left[ ^H\;_V\right] {}^\alpha {}_\beta
{}=-S_N\left[ ^H{}_V\right] {}^\alpha {}_\beta {}+D_{V\beta }N^{H\alpha
}+h_H^T{}_\beta {}^\alpha {}_\sigma {}{}N^{H\sigma }-2\omega _V{}^\alpha
{}_{\beta \sigma }{}N^{V\sigma }
\end{equation}
\begin{equation}
\label{Lie.H.VH.comp}\pounds _NH\left[ ^V\;_H\right] {}^\alpha {}_\beta
{}=S_N\left[ ^V{}_H\right] {}^\alpha {}_\beta {}-D_{H\beta }N^{V\alpha
}-h_V^T{}_\beta {}^\alpha {}_\sigma {}N^{V\sigma }+2\omega _H{}^\alpha
{}_{\beta \sigma }{}N^{H\sigma }
\end{equation}
where the spacetime torsion terms are components of the assembly
\begin{equation}
\label{SN.def}S_N\left[ \,\right] {}^{\left\langle \alpha \right\rangle
}{}_{\left\langle \beta \right\rangle }=S\left[ \,\right] {}^{\left\langle
\alpha \right\rangle }{}_{\left\langle \beta \right\rangle \left\langle
\sigma \right\rangle }{}N^{\left\langle \sigma \right\rangle }.
\end{equation}

\subsubsection{The metric tensor}

These techniques also yield an expression for the assembly of projections of
the Lie derivative of the metric. Express the Lie derivative in terms of a
covariant derivative and decompose the result into its restricted parts.
Just as was the case for the Lie derivative of the projection tensor field,
the result can be simplified in two ways. In terms of assemblies, the result
is
\begin{equation}
\label{Lie.g.asm}
\begin{array}{c}
\pounds _Ng\left[ \,\right] ^{\left\langle \mu \right\rangle \left\langle
\nu \right\rangle }=\left[ 2g^{\left\langle \rho \right\rangle (\left\langle
\mu \right\rangle }S^{\left\langle \nu \right\rangle )}{}_{\left\langle \rho
\right\rangle \left\langle \delta \right\rangle }-Q{}^{\left\langle \mu
\right\rangle \left\langle \nu \right\rangle }{}_{\left\langle \delta
\right\rangle }\right] N^{\left\langle \delta \right\rangle } \\
-g^{\left\langle \rho \right\rangle \left\langle \nu \right\rangle
}D_{\left\langle \rho \right\rangle }N^{\left\langle \mu \right\rangle
}-g^{\left\langle \mu \right\rangle \left\langle \rho \right\rangle
}D_{\left\langle \rho \right\rangle }N^{\left\langle \nu \right\rangle }.
\end{array}
\end{equation}
In terms of restricted objects, the projections of the Lie derivative are
\begin{equation}
\label{Killing.hh}
\begin{array}{c}
\pounds _Ng\left[ ^{HH}\,\right] ^{\mu \nu }=-Q_N\left[ ^{HH}\,\right]
{}^{\mu \nu } \\
-2g^{HH\rho (\nu }D_{H\rho }N^{H\mu )}-2g^{HV\rho (\mu }D_{V\rho }N^{H\nu )}
\\
+\left[ g^{HH\mu \rho }h_H^T{}_\sigma {}^\nu {}_\rho {}+g^{HH\rho \nu
}h_H^T{}_\sigma {}^\mu {}_\rho {}-g^{HV\mu \rho }h_V{}^\nu {}_{\sigma \rho
}{}-g^{VH\rho \nu }h_V{}^\mu {}_{\sigma \rho }{}\right] N^{V\sigma }
\end{array}
\end{equation}
\begin{equation}
\label{Killing.hv}
\begin{array}{c}
\pounds _Ng\left[ \,^{HV}\right] ^{\mu \nu }=-Q_N\left[ ^{HV}\,\right]
{}^{\mu \nu } \\
-\left( g^{HV\rho \nu }D_{H\rho }+g^{VV\rho \nu }D_{V\rho }\right) N^{H\mu
}-\left( g^{HH\mu \rho }D_{H\rho }+g^{HV\mu \rho }D_{V\rho }\right) N^{V\nu
} \\
-\left[ g^{HH\mu \rho }h_H{}^\nu {}_{\sigma \rho }{}-g^{HV\mu \rho
}h_V^T{}_\sigma {}^\nu {}_\rho {}\right] N^{H\sigma }+\left[ -g^{VV\rho \nu
}h_V{}^\mu {}_{\sigma \rho }{}+g^{HV\rho \nu }h_H^T{}_\sigma {}^\mu {}_\rho
{}\right] N^{V\sigma }
\end{array}
\end{equation}
and the complements of these expressions. Here, the terms which normally
vanish in Riemannian geometries have been collected into the assembly
\begin{equation}
\label{QN.def}Q_N\left[ \,\right] {}^{\left\langle \mu \right\rangle
\left\langle \nu \right\rangle }{}=\left( Q\left[ \,\right] {}^{\left\langle
\mu \right\rangle \left\langle \nu \right\rangle }{}_{\left\langle \delta
\right\rangle }-g^{\left\langle \mu \right\rangle \left\langle \rho
\right\rangle }S\left[ \,\right] {}^{\left\langle \nu \right\rangle
}{}_{\left\langle \rho \right\rangle \left\langle \delta \right\rangle
}-g^{\left\langle \rho \right\rangle \left\langle \nu \right\rangle }S\left[
\,\right] {}^{\left\langle \mu \right\rangle }{}_{\left\langle \rho
\right\rangle \left\langle \delta \right\rangle }\right) N^{\left\langle
\delta \right\rangle }.
\end{equation}

\subsubsection{Restricted metric tensors}

Consider each projection $g^{HH},g^{HV},g^{VH},g^{VV}$ of the metric tensor
as a separate restricted tensor field, compute the Lie derivatives of these
tensor fields, and project the results in all possible ways. In each case,
the result can be represented in the form of a projection assembly. The
resulting assemblies can be found easily by applying the product rule and
Eqs. (\ref{Lie.g.asm}) and (\ref{Lie.H.asm}) to expressions such as
$$
g^{HH}{}^{\left\langle \alpha \right\rangle \left\langle \beta \right\rangle
}=H^{\left\langle \alpha \right\rangle }{}_{\left\langle \rho \right\rangle
}{}g^{\left\langle \rho \right\rangle \left\langle \sigma \right\rangle
}{}H^{\left\langle \beta \right\rangle }{}_{\left\langle \sigma
\right\rangle }{}
$$
in order to obtain
\begin{equation}
\label{LN-gHH}
\begin{array}{c}
\pounds _Ng^{HH}{}^{\left\langle \alpha \right\rangle \left\langle \beta
\right\rangle }=-2g^{\left\langle H\rho \right\rangle (\left\langle H\beta
\right\rangle }D_{\left\langle H\rho \right\rangle }N^{\left\langle \alpha
\right\rangle )} \\
+\left( 2g^{\left\langle H\rho \right\rangle (\left\langle H\beta
\right\rangle }S{}^{\left\langle \alpha \right\rangle )}{}_{\left\langle
H\rho \right\rangle \left\langle \sigma \right\rangle
}{}{}-Q{}^{\left\langle H\alpha \right\rangle \left\langle H\beta
\right\rangle }{}_{\left\langle \sigma \right\rangle }\right)
N^{\left\langle \sigma \right\rangle }
\end{array}
\end{equation}
\begin{equation}
\label{LN-gHV}
\begin{array}{c}
\pounds _Ng^{HV}{}^{\left\langle \alpha \right\rangle \left\langle \beta
\right\rangle }=-g^{\left\langle H\rho \right\rangle \left\langle V\beta
\right\rangle }D_{\left\langle H\rho \right\rangle }N^{\left\langle \alpha
\right\rangle } \\
-g^{\left\langle H\alpha \right\rangle \left\langle V\rho \right\rangle
}D_{\left\langle V\rho \right\rangle }N^{\left\langle H\beta \right\rangle
}-g^{\left\langle H\alpha \right\rangle \left\langle H\rho \right\rangle
}D_{\left\langle H\rho \right\rangle }N^{\left\langle V\beta \right\rangle }
\\
+\left( g^{\left\langle H\rho \right\rangle \left\langle V\beta
\right\rangle }S{}^{\left\langle \alpha \right\rangle }{}_{\left\langle
H\rho \right\rangle \left\langle \sigma \right\rangle }{}+g^{\left\langle
H\alpha \right\rangle \left\langle V\rho \right\rangle }S^{\left\langle
\beta \right\rangle }{}_{\left\langle V\rho \right\rangle \left\langle
\sigma \right\rangle }-Q{}^{\left\langle H\alpha \right\rangle \left\langle
V\beta \right\rangle }{}_{\left\langle \sigma \right\rangle }\right)
N^{\left\langle \sigma \right\rangle }
\end{array}
\end{equation}
In terms of restricted objects, and the normally vanishing assemblies $Q_N$
and $S_N$ defined by Eqs. \ref{QN.def} and \ref{SN.def}, the projections of
these Lie derivatives are given by
\begin{equation}
\label{LN-gHHHH}
\begin{array}{c}
L_Ng^{HH}{}^{\alpha \beta }=-Q_N\left[ ^{HH}\right] {}^{\alpha \beta }{} \\
-2g^{HH\rho (\alpha }D_{H\rho }N^{H\beta )}+2g^{VH}{}^{\rho (\beta
}{}h_H^T{}_\rho {}^{\alpha )}{}_\sigma {}N^{H\sigma } \\
+2\left( g^{HH\rho (\alpha }h_H^T{}_\sigma {}^{\beta )}{}_\rho
{}-g^{VH}{}^{\rho (\alpha }{}h_V{}^{\beta )}{}_{\rho \sigma }{}\right)
N^{V\sigma }{}
\end{array}
\end{equation}
\begin{equation}
\label{LN-gHHHV}
\begin{array}{c}
\pounds _Ng^{HH}{}\left[ \,^{HV}\right] {}^{\alpha \beta }=g^{HH\rho \alpha
}S_N\left[ ^V{}_H\right] {}^\beta {}_{\rho \sigma }{} \\
-g^{HH\rho \alpha }D_{H\rho }N^{V\beta }+2g^{HH\rho \alpha }h_H{}^\beta
{}_{\left[ \rho \sigma \right] }{}N^{H\sigma }-g^{HH\rho \alpha
}h_V^T{}_\rho {}^\beta {}_\sigma {}N^{V\sigma }
\end{array}
\end{equation}
\begin{equation}
\label{LN-gHHVV}\pounds _Ng^{HH}{}\left[ \,^{VV}\right] {}^{\alpha \beta }=0
\end{equation}
\begin{equation}
\label{LN-gHVHV}
\begin{array}{c}
L_Ng^{HV}{}^{\alpha \beta }=-Q_N\left[ ^{HV}\right] {}^{\alpha \beta }{} \\
-g^{HV\rho \beta }D_{H\rho }N^{H\alpha }-g^{HH\alpha \rho }D_{H\rho
}N^{V\beta } \\
+\left( g^{HV\alpha \rho }h_V^T{}_\sigma {}^\beta {}_\rho
{}-g^{HH}{}^{\alpha \rho }{}h_H{}^\beta {}_{\rho \sigma }{}+g^{VV}{}^{\rho
\beta }{}h_H^T{}_\rho {}^\alpha {}_\sigma {}\right) N^{H\sigma } \\
+\left( g^{HV\rho \beta }h_H^T{}_\sigma {}^\alpha {}_\rho {}-g^{VV}{}^{\rho
\beta }{}h_V{}^\alpha {}_{\rho \sigma }{}+g^{HH}{}^{\alpha \rho
}{}h_V^T{}_\rho {}^\beta {}_\sigma {}\right) N^{V\sigma }
\end{array}
\end{equation}
\begin{equation}
\label{LN-gHVHH}
\begin{array}{c}
\pounds _Ng^{HV}{}\left[ ^{HH}\right] {}^{\alpha \beta }=g^{HV\alpha \rho
}S_N\left[ ^H{}_V\right] {}^\beta {}_{\rho \sigma }{} \\
-g^{HV\alpha \rho }\left( D_{V\rho }N^{H\beta }+h_H^T{}_\rho {}^\beta
{}_\sigma {}N^{H\sigma }-2h_V{}^\beta {}_{\left[ \rho \sigma \right]
}{}N^{V\sigma }\right)
\end{array}
\end{equation}
The remaining projections can be obtained by taking the complements of these
and by using the symmetry of the metric tensor.

\subsubsection{Generalized area change}

An important consequence of Eq. \ref{LN-gHHHH} is a formula for the
evolution of a generalized area element. Suppose that the dimension of $HT_P$
is $s$. The area element on surfaces tangent to the subspace $HT_P$ is just
the unit $s$-form $\alpha $ in $H^{*}\hat T_P$. Choose a vector field $%
N^\alpha $ in $VT_P$ and an $s$-form field $\sigma $ which is propagated
along the integral curves of $N^\alpha $ by Lie-dragging so that it solves $%
\pounds _N\sigma =0$. So long as the chosen Lie-dragged $s$-form $\sigma $
obeys%
$$
H^{*}\sigma \neq 0,\qquad g^{-1}\left( H^{*}\sigma ,H^{*}\sigma \right) \neq
0,
$$
the $H$-area element $\alpha $ can be constructed from $\sigma .$ If the
subspace $HT_P$ is spacelike $\left( k=1\right) $ or timelike $\left(
k=-1\right) $, then%
$$
\alpha =\left[ kg^{-1}\left( H^{*}\sigma ,H^{*}\sigma \right) \right]
^{-1/2}H^{*}\sigma .
$$
Because the restricted Lie derivative obeys $L_NH=0$ and the field $\sigma $
has been defined by Lie-dragging, it is easy to see that $L_N\left(
H^{*}\sigma \right) =0$. The restricted Lie derivative of the $H$-area
element is then%
$$
L_N\alpha =-\frac 12\left[ g^{-1}\left( H^{*}\sigma ,H^{*}\sigma \right)
\right] ^{-1}\left\{ L_Ng^{-1}\right\} \left( H^{*}\sigma ,H^{*}\sigma
\right) \alpha
$$
Expand the forms in an adapted frame and find that $g^{-1}\left( H^{*}\sigma
,H^{*}\sigma \right) $ is just the determinant of the matrix of coefficients
$g^{HH}{}^{ab}$ and, from the usual formula for the derivative of a
determinant,

$$
L_N\alpha =-\frac 12g_{HH}{}_{\alpha \beta }L_Ng^{HH}{}^{\alpha \beta
}\,\alpha .
$$
Eq. \ref{LN-gHHHH} now gives the result
\begin{equation}
\label{gen-area-change}g_{HH}{}_{\alpha \beta }\left( g^{HH\rho (\alpha
}h_H^T{}_\sigma {}^{\beta )}{}_\rho {}-g^{VH}{}^{\rho (\alpha
}{}h_V{}^{\beta )}{}_{\rho \sigma }{}\right) N^\sigma {}\,\alpha
=-\,L_N\alpha
\end{equation}

Notice that, in general, $g^{HH\rho \alpha }$ is not the matrix inverse of $%
g_{HH}{}_{\alpha \beta }$. By projecting the identity $g_{\alpha \beta
}g^{\rho \alpha }=\delta _\beta ^\rho $, one finds that the above equation
takes the form%
$$
N^\sigma {}\theta _H^T{}_\sigma {}\alpha -\left( g_{VH}{}_{\alpha \beta
}g^{HV}{}^{\rho \alpha }h_H^T{}_\sigma {}^\beta {}_\rho {}+g_{HH}{}_{\alpha
\beta }g^{VH}{}^{\rho \alpha }{}h_V{}^\beta {}_{\rho \sigma }{}\right)
N^\sigma \,\alpha =-\,L_N\alpha
$$
For a normal projection-tensor field, $g^{HV}{}^{\rho \alpha }=0$ and the
familiar relation
\begin{equation}
\label{area-change}N^\sigma {}\theta _H^T{}_\sigma {}\,\alpha =-\,L_N\alpha
\end{equation}
between the divergence and the rate of change of the projected area element
is obtained.

\subsection{Rule for differentiating restricted tensor fields}

\subsubsection{Restricted vector and form fields}

Start with the expressions%
$$
\begin{array}{c}
v^{\left\langle \alpha \right\rangle }=H^{\left\langle \alpha \right\rangle
}{}_{\left\langle \rho \right\rangle }v^{\left\langle \rho \right\rangle
}\qquad
\text{if}\qquad v\in HT_P \\ v^{\left\langle \alpha \right\rangle
}=V^{\left\langle \alpha \right\rangle }{}_{\left\langle \rho \right\rangle
}v^{\left\langle \rho \right\rangle }\qquad \text{if}\qquad v\in VT_P
\end{array}
$$
Differentiate these expressions%
$$
\pounds _Nv^{\left\langle \alpha \right\rangle }=\left\{
\begin{array}{rrr}
\left( \pounds _NH^{\left\langle \alpha \right\rangle }{}_{\left\langle \rho
\right\rangle }\right) v^{\left\langle \rho \right\rangle }+H^{\left\langle
\alpha \right\rangle }{}_{\left\langle \rho \right\rangle }\pounds
_Nv^{\left\langle \rho \right\rangle } & \text{if} & v\in HT_P \\
-\left( \pounds _NH^{\left\langle \alpha \right\rangle }{}_{\left\langle
\rho \right\rangle }\right) v^{\left\langle \rho \right\rangle
}+V^{\left\langle \alpha \right\rangle }{}_{\left\langle \rho \right\rangle
}\pounds _Nv^{\left\langle \rho \right\rangle } & \text{if} & v\in VT_P
\end{array}
\right.
$$
and notice that they are summarized by%
$$
\pounds _Nv^{\left\langle \alpha \right\rangle }=\left( \pounds
_NH^{\left\langle \alpha \right\rangle }{}_{\left\langle \sigma
\right\rangle }\right) C{}^{\left\langle \sigma \right\rangle
}{}_{\left\langle \rho \right\rangle }{}v^{\left\langle \rho \right\rangle
}+L_Nv^{\left\langle \alpha \right\rangle }
$$

Similarly, differentiate the projection identities satisfied by restricted
forms and obtain%
$$
\pounds _N\varphi _{\left\langle \beta \right\rangle }=\varphi
_{\left\langle \rho \right\rangle }C{}^{\left\langle \rho \right\rangle
}{}_{\left\langle \sigma \right\rangle }\left( \pounds _NH^{\left\langle
\sigma \right\rangle }{}_{\left\langle \beta \right\rangle }\right)
+L_N\varphi _{\left\langle \beta \right\rangle }
$$

\subsubsection{The Lie-derivative correction assembly}

Evidently, the assembly%
$$
\ell _N^{\prime }{}^{\left\langle \alpha \right\rangle }{}_{\left\langle
\beta \right\rangle }=C{}^{\left\langle \alpha \right\rangle
}{}_{\left\langle \sigma \right\rangle }\left( \pounds _NH^{\left\langle
\sigma \right\rangle }{}_{\left\langle \beta \right\rangle }\right) =-\left(
\pounds _NH^{\left\langle \alpha \right\rangle }{}_{\left\langle \sigma
\right\rangle }\right) C{}^{\left\langle \sigma \right\rangle
}{}_{\left\langle \beta \right\rangle }{}
$$
plays the same role in the projection decomposition of Lie derivatives as
the quantity $\nabla _\rho ^{\prime }N^\delta $ plays in their covariant
derivative expressions. From Eq. \ref{Lie.HC.asm} the general expression for
this assembly is
\begin{equation}
\label{l.C.asm}
\begin{array}{c}
\ell _N^{\prime }{}^{\left\langle \alpha \right\rangle }{}_{\left\langle
\rho \right\rangle }=
\frac 12\left( C^{\left\langle \tau \right\rangle }{}_{\left\langle \sigma
\right\rangle }{}D_{\left\langle \tau \right\rangle }N^{\left\langle \alpha
\right\rangle }-C^{\left\langle \alpha \right\rangle }{}_{\left\langle \tau
\right\rangle }{}D_{\left\langle \sigma \right\rangle }N^{\left\langle \tau
\right\rangle }\right) C{}^{\left\langle \sigma \right\rangle
}{}_{\left\langle \rho \right\rangle }{} \\ +\frac 12N^{\left\langle
\varsigma \right\rangle }\left( C^{\left\langle \alpha \right\rangle
}{}_{\left\langle \tau \right\rangle }{}S{}^{\left\langle \tau \right\rangle
}{}_{\left\langle \sigma \right\rangle \left\langle \varsigma \right\rangle
}{}-C^{\left\langle \tau \right\rangle }{}_{\left\langle \sigma
\right\rangle }{}S{}^{\left\langle \alpha \right\rangle }{}_{\left\langle
\tau \right\rangle \left\langle \varsigma \right\rangle }{}\right)
C{}^{\left\langle \sigma \right\rangle }{}_{\left\langle \rho \right\rangle
}{}.
\end{array}
\end{equation}

This expression is a good example of a case where the assembly notation is
less compact and transparent than writing out all of its components. Define
the {\em Lie derivative correction assembly }%
$$
\ell _N{}^{\left\langle \alpha \right\rangle }{}_{\left\langle \rho
\right\rangle }=\ell _N^{\prime }{}^{\left\langle \alpha \right\rangle
}{}_{\left\langle \rho \right\rangle }+S_N\left[ \,\right] {}^{\left\langle
\alpha \right\rangle }{}_{\left\langle \rho \right\rangle }
$$
which plays a role analogous to the one played by $\nabla _\rho N^\delta $
in the rendering of Lie derivatives in terms of covariant derivatives. From
Eqs. \ref{Lie.H.HV.comp} and \ref{Lie.H.VH.comp} this assembly can be
expressed quite compactly by%
$$
\ell _N{}^{\left\langle H\alpha \right\rangle }{}_{\left\langle H\beta
\right\rangle }=0,
$$
$$
\ell _N{}^{\left\langle V\alpha \right\rangle }{}_{\left\langle H\beta
\right\rangle }=D_{H\beta }N^{V\alpha }+h_V^T{}_\beta {}^\alpha {}_\sigma
{}N^{V\sigma }-2\omega _H{}^\alpha {}_{\beta \sigma }{}N^{H\sigma },
$$
and the complements of these expressions.

\subsubsection{Restricted tensor fields of arbitrary rank}

Consider an arbitrary-rank, restricted tensor-field $M^{\delta _1\delta
_2\ldots \delta _n}{}_{\alpha _1\alpha _2\ldots \alpha _n}.$ Such a tensor
field obeys a projection identity given by Eq. \ref{OM.identity}. The Lie
derivative of such a tensor field can be expressed in terms of restricted
tensors by taking the Lie derivative of this identity with the result.
\begin{equation}
\label{Lie.tensor.prjct}
\begin{array}{c}
\left( \pounds _NM\right) ^{\left\langle \delta _1\right\rangle \ldots
\left\langle \delta _n\right\rangle }{}_{\left\langle \alpha _1\right\rangle
\ldots \left\langle \alpha _n\right\rangle }=L_NM^{\left\langle \delta
_1\right\rangle \ldots \left\langle \delta _n\right\rangle }{}_{\left\langle
\alpha _1\right\rangle \ldots \left\langle \alpha _n\right\rangle } \\
-M^{\left\langle \rho \right\rangle \ldots \left\langle \delta
_n\right\rangle }{}_{\left\langle \alpha _1\right\rangle \ldots \left\langle
\alpha _n\right\rangle }\ell _N^{\prime }{}^{\left\langle \delta
_1\right\rangle }{}_{\left\langle \rho \right\rangle }{}-\ldots
-M^{\left\langle \delta _1\right\rangle \ldots \left\langle \rho
\right\rangle }{}_{\left\langle \alpha _1\right\rangle \ldots \left\langle
\alpha _n\right\rangle }\ell _N^{\prime }{}^{\left\langle \delta
_n\right\rangle }{}_{\left\langle \rho \right\rangle }{} \\
+M^{\left\langle \delta _1\right\rangle \ldots \left\langle \delta
_n\right\rangle }{}_{\left\langle \rho \right\rangle \ldots \left\langle
\alpha _n\right\rangle }\ell _N^{\prime }{}^{\left\langle \rho \right\rangle
}{}_{\left\langle \alpha _1\right\rangle }{}+\ldots +M^{\left\langle \delta
_1\right\rangle \ldots \left\langle \delta _n\right\rangle }{}_{\left\langle
\alpha _1\right\rangle \ldots \left\langle \rho \right\rangle }\ell
_N^{\prime }{}^{\left\langle \rho \right\rangle }{}_{\left\langle \alpha
_n\right\rangle }{}.
\end{array}
\end{equation}

\subsubsection{Decomposing derivatives of unrestricted tensors}

The preceding result may be used to express the projections of the Lie
derivative of an unrestricted tensor $m$ in terms of its projections. Simply
express the tensor $m$ as the sum of its projections, sum the preceding
expression over all projection labels, and project the result on all free
indexes. The resulting expression for $\left( \pounds _Nm\right) \left[
\,\right] ^{\left\langle \delta _1\right\rangle \ldots \left\langle \delta
_n\right\rangle }{}_{\left\langle \alpha _1\right\rangle \ldots \left\langle
\alpha _n\right\rangle }$ looks exactly like Eq. \ref{Lie.tensor.prjct} with
$M$ replaced by $m.$

\section{Projections onto isometry group orbits}
\label{orbits}

\subsection{Basics}

\subsubsection{Background}

The study of metric-tensor symmetries is arguably the oldest topic in the
field of general relativity. Most results in this area have been discovered
and re-discovered, formulated, and re-formulated many times as fashions in
notation have changed. This section is yet another chapter in this history
of reformulation. The tilted projection tensor formulation is new and has
some new insights to offer but, as far as I can tell, the particular
properties which I will be discussing are not new. I apologize in advance to
the very large number of colleagues whose work I have surely failed to cite.

\subsubsection{Killing vectors}

A vector field $N$ on a manifold $M$ generates an {\em isometry} or,
equivalently, a {\em motion of the metric tensor}, if the corresponding Lie
derivative of the metric tensor vanishes everywhere on $M$.\cite{Killing-def}
In the most general case, the resulting condition on $N$ is%
$$
\pounds _Ng^{\alpha \beta }=-Q{}^{\alpha \beta }{}_\delta N^\delta -2g^{\rho
(\alpha }\nabla _\rho ^{\prime }N^{\beta )}=0.
$$
A vector field which satisfies this condition everywhere on $M$ is called a
{\em Killing vector field}. Within this section, {\em I will be assuming a
metric-compatible, torsion-free connection} so that Killing vectors obey the
familiar form of Killing's equation.%
$$
g^{\rho (\alpha }\nabla _\rho N^{\beta )}=0
$$

\subsubsection{Group orbits}

Each Killing vector field generates a map of the manifold into itself. The
set of all points $O_P$ which can be reached by such maps, starting from a
given point $P\in M$ is called the {\em isometry group orbit} through the
point $P$. If there are $n$ Killing vector fields then each orbit is a
submanifold of dimension $r\leq n$. When the dimension $r$ of a group orbit
is less than the dimensionality $n$ of the isometry group, the Killing
vectors cannot all be linearly independent everywhere on that orbit --- At
each point $P$ of the orbit, some linear combination $\Sigma _PN$ of Killing
vectors must vanish. The point $P$ is then a fixed point of the motion
generated by the combination $\Sigma _PN$. Spherical symmetry, with $n=3$, $%
r=2$ is the best known example of this situation.

\subsubsection{Isotropy}

Isotropies constrain the direction of invariant vector fields. If $P$ is a
fixed point of a subgroup $I_P$ of a group of motions $G$ then $I_P$ induces
mappings on the vector space $HT_P$ which is tangent to the group orbit
through $P$. The group $G$ is said to have an isotropic orbit at $P$ if the
subgroup $I_P$ induces mappings which connect any two directions (i.e. rays)
in $HT_P$. In other words, there are {\em no preferred directions} on an
isotropic orbit. The result that will be needed here is a very simple one:
If $G$ has an isotropic orbit, and a vector-field is tangent to the orbit
and invariant under $G$ then the vector-field vanishes everywhere on the
orbit. This result can be used to obtain powerful constraints on an
isotropic geometry by the technique of constructing group-invariant vector
fields tangent to the group orbits and then setting those vector fields
equal to zero.

\subsection{The general case}

\subsubsection{Arbitrary orbit projections}

Now consider a geometry with an isometry group and a projection tensor field
$H$ which projects vectors into the tangent spaces to group orbits. For any
Killing vector field $\xi $ one then has the relations $H\xi =\xi $ or,
equivalently, $\xi ^{H\alpha }=\xi ^\alpha $ and $\xi ^{V\alpha }=0$ as well
as Killing's equation $\pounds _\xi g^{\mu \nu }=0$. Equations. \ref
{Killing.hh}, and \ref{Killing.hv} for the projections of the Lie derivative
of the metric tensor then require
\begin{equation}
\label{int.killling.eqn}2g^{HH\rho (\nu }D_{H\rho }\xi ^{\mu )}+2g^{HV\rho
(\mu }D_{V\rho }\xi ^{\nu )}=0
\end{equation}
and
\begin{equation}
\label{ext.killing.eqn}
\begin{array}{c}
\left( g^{HV\rho \nu }D_{H\rho }+g^{VV\rho \nu }D_{V\rho }\right) \xi ^\mu
\\
+\left[ g^{HH\mu \rho }h_H{}^\nu {}_{\sigma \rho }{}-g^{HV\mu \rho
}h_V^T{}_\sigma {}^\nu {}_\rho {}\right] \xi ^\sigma =0
\end{array}
\end{equation}
while the complement of Eq. \ref{Killing.hh} requires
\begin{equation}
\label{h.killing.eqn}\left[ g^{VV\mu \rho }h_V^T{}_\sigma {}^\nu {}_\rho
{}+g^{VV\rho \nu }h_V^T{}_\sigma {}^\mu {}_\rho {}-g^{VH\mu \rho }h_H{}^\nu
{}_{\sigma \rho }{}-g^{HV\rho \nu }h_H{}^\mu {}_{\sigma \rho }{}\right] \xi
^\sigma =0.
\end{equation}
The complement of Eq. \ref{Killing.hv} simply requires Eq. \ref
{ext.killing.eqn} again.

Because the Killing vectors span the tangent space $HT_P$ to the group
orbit, Eq. \ref{h.killing.eqn} implies a restriction on the projection
curvatures which can be written in the form
\begin{equation}
\label{h.TV.killing.eqn}g^{VV\rho (\nu }h_V^T{}_\sigma {}^{\mu )}{}_\rho
{}=g^{HV\rho (\nu }h_H{}^{\mu )}{}_{\sigma \rho }{}.
\end{equation}
This restriction holds for any projection onto any group orbit no matter how
the projection is tilted. In addition to this restriction, the fact that the
projection is onto the tangent spaces to a surface requires (see Ref. 1)
\begin{equation}
\label{hH.twist.killing.eqn}h_H{}^\mu {}_{\sigma \rho }{}=h_H{}^\mu {}_{\rho
\sigma }{}.
\end{equation}

Equations \ref{int.killling.eqn} and \ref{ext.killing.eqn} imply
restrictions on some of the connection coefficients. Because the Killing
vectors $\left\{ \xi _a\right\} $ span the tangent space to the orbit, the
coefficients defined by%
$$
\Gamma _H^H{}^\mu {}_{a\delta }=D_{H\delta }\xi _a^\mu ,\qquad \Gamma
_V^H{}^\mu {}_{a\delta }=D_{V\delta }\xi _a^\mu
$$
contain all of the information needed to evaluate restricted covariant
derivatives of tensor fields on an orbit. The restrictions on these
coefficients are
\begin{equation}
\label{group-cnctn}2g^{HH}{}^{\rho (\nu }{}\Gamma _H^H{}^{\mu )}{}_{a\rho
}+2g^{HV}{}^{\rho (\nu }{}\Gamma _V^H{}^{\mu )}{}_{a\rho }=0
\end{equation}
\begin{equation}
\label{Fermi-cnctn}g^{HV}{}^{\rho \nu }{}\Gamma _H^H{}^\mu {}_{a\rho
}+g^{VV}{}^{\rho \nu }{}\Gamma _V^H{}^\mu {}_{a\rho }+\left[ g^{HH\mu \rho
}h_H{}^\nu {}_{\sigma \rho }{}-g^{HV\mu \rho }h_V^T{}_\sigma {}^\nu {}_\rho
{}\right] \xi _a^\sigma =0
\end{equation}
Equation \ref{group-cnctn} expresses the compatibility between the
restricted connection induced on an orbit and the isometry group which is
transitive on that orbit. Equation \ref{Fermi-cnctn} ensures that the Fermi
derivative ($D_V$ in this case) is compatible with the isometry group.

\subsubsection{Group invariant orbit projections}

The projection tensor field can be specialized further. At least within an
open set $\Omega $ which contains a non-degenerate group orbit one can find
a reference surface $S_{\text{R}}$ which intersects each group orbit in $%
\Omega $ exactly once. At each point $P$ on $S_{\text{R}}$ let $HT_P$ be the
tangent space to the group orbit through the point $P$ and choose $VT_P$
arbitrarily, thus specifying the tensor $H$ on $S_{\text{R}}$. Now use the
Killing vector fields to Lie-drag this projection-tensor throughout $\Omega $%
. The result of this construction is a group-invariant projection onto the
group orbits. Throughout the rest of this paper, it will be assumed that
projections are group-invariant.

{}From Eq.\ref{Lie.H.HV.comp} as well as the complement of Eq. \ref
{Lie.H.VH.comp}, the vanishing Lie derivative of $H$ then requires each
Killing vector to obey the restriction:
\begin{equation}
\label{Dv.eqn}D_{V\beta }\xi ^\alpha =-h_H^T{}_\beta {}^\alpha {}_\sigma
{}{}\xi ^\sigma .
\end{equation}
For a set $\left\{ \xi _a\right\} $ of Killing vectors which span the
tangent space to the orbit, this restriction provides a direct formula for
the mixed or Fermi-derivative connection coefficients
\begin{equation}
\label{Fermi-coeff}\Gamma _V^H{}^\alpha {}_{a\beta }{}=-h_H^T{}_\beta
{}^\alpha {}_\sigma {}{}\xi _a^\sigma
\end{equation}
and thus describes how to take the restricted derivative $D_V$ of
orbit-tangent tensor fields. Equation \ref{Dv.eqn}, combined with the
definition of the restricted curvature tensor, yields the result
\begin{equation}
\label{R-HVV}R{}_{VV}^H{}_\tau {}^\alpha {}_{\beta \gamma }=2D_{V[\beta
}h_H^T{}_{\gamma ]}{}^\alpha {}_\tau {}+h_H^T{}_{[\beta }{}^\alpha
{}_{\left| \sigma \right| }{}h_H^T{}_{\gamma ]}{}^\sigma {}_\tau {}.
\end{equation}

\subsubsection{Group-invariant evolutions}

A {\em group-invariant vector-field} $N$ is a vector field which commutes
with all of the Killing vectors. I will define a {\em group-invariant
evolution of the orbit O} to be a finite set of group-invariant vector
fields $N\left[ K\right] $ such that the vectors $VN\left[ K\right] \left(
P\right) $ span the space $VT_P$ for every point $P$ on O. It is easy to see
that non-null, orientable isometry orbits of codimension = 1 always have
group-invariant evolutions --- the unit normal vector fields. In a generic
spacetime with a group-invariant projection, one might construct a complete
set of group-invariant vectors algebraically from the projected Riemann
tensor $R_{VV}^V{}_\alpha {}^\rho {}_{\mu \nu }$ and its restricted
derivatives. For example, in codimension = 2, the vector $N^\alpha
=g^{VV}{}^{\alpha \delta }D_{V\delta }{}R_{VV}^V$ and a unit vector
orthogonal to it in $VT_P$ will provide a group-invariant evolution for
those group orbits where $D_{V\delta }{}R_{VV}^V$ is not null. In general,
however, such constructions are difficult to carry out and not particularly
instructive. In what follows, I will simply restrict my attention to group
orbits which have group-invariant evolutions.

\subsubsection{The divergence-orbit-area relation}

For compact orbits, it is possible to extend Eq. \ref{area-change} for the
evolution of an area-element by integrating it over each orbit.
\begin{equation}
\label{int-area-change}\int_{O_P}\alpha N^\alpha \theta _H^T{}_\alpha
{}=-\int_{O_P}L_N\alpha
\end{equation}
The assumption that this orbit has a group-invariant evolution lets us
choose $N^\alpha $ to be group-invariant. Since $\theta {}_\alpha {}$is
constructed from the projection tensor field $H$ and the connection, it will
be group-invariant if $H$ is group-invariant. Thus, the scalar $N^\alpha
\theta {}_\alpha {}$ is constant on the orbit and can be moved outside the
integral. Because $N$ commutes with the group generators, the Lie derivative
on the right-hand side of the equation can be replaced by a derivative of
the orbit-area function defined at each point $P$ by%
$$
a\left( P\right) =\int_{O_P}\alpha .
$$
Eq. \ref{int-area-change} then becomes%
$$
N^\alpha \theta _H^T{}_\alpha {}a\left( P\right) =-L_Na\left( P\right)
=-N\cdot da=-N\cdot V^{*}da=-N^\alpha D_{V\alpha }a.
$$
Because $N$ is part of a group-invariant evolution which spans $VT_P$, this
result implies $\theta _H^T{}_\alpha {}a=-D_{V\alpha }a$ so that the
divergence is the gradient of a potential.
\begin{equation}
\label{div-grad}\theta _H^T{}_\alpha {}=-D_{V\alpha }\left( \ln a\right) .
\end{equation}

When the group orbits are not compact, this same argument can often be
applied to a finite part $A$ of an orbit which is mapped onto nearby orbits
by the group-invariant evolution. The main complication which can occur is
that the projection $V$ and therefore the group-invariant evolution which
spans it may not be surface-forming. In that case the vector fields in the
evolution might not map $A$ into the same subsets of neighboring orbits.
When the $V$-twist tensor $\omega _V{}^{\,\rho }{}_{\nu \gamma }{}$ is zero,
$V$ is surface-forming and Eq. \ref{div-grad} still holds. Because the
result depends on the logarithmic derivative of the area, it is not affected
by the size of the chosen subset $A$.

\subsection{Normal projections onto group orbits}

\subsubsection{Normality}

Although this sequence of papers removes the assumption that projection
tensor fields are normal and discusses the geometry of arbitrarily tilted
projections, it is important to understand just how powerful the normality
condition is. At the level of the metric tensor, the assumption has a simple
statement:
\begin{equation}
\label{normality}g^{HV\alpha \beta }=0.
\end{equation}
{}From this normality assumption with the further assumption%
$$
Q{}^{\mu \nu }{}_\delta {}=0
$$
of metric compatibility for the full connection and the metricity
decompositions (Eqs. \ref{intr.metricity}, \ref{fermi.metricity}, \ref
{hT-h.relation}, and \ref{hTV-h.relation}) come three remarkable results:
(1) From Eq. \ref{intr.metricity} and its complement comes the compatibility
of the restricted derivatives $D_{H\delta }$ and $D_{V\delta }$ with the
intrinsic metrics $g^{HH}{}^{\alpha \beta }$ and $g^{VV}{}^{\alpha \beta }$
on the corresponding subspaces
$$
Q_H^{HH}{}^{\mu \nu }{}_\delta {}=0,\qquad Q_V^{VV}{}^{\mu \nu }{}_\delta
{}=0.
$$
(2) From Eq. \ref{fermi.metricity} and its complement come the vanishing
Fermi derivatives of these intrinsic metrics%
$$
Q_V^{HH}{}^{\mu \nu }{}_\delta {}=-D_V{}_\delta {}g^{HH}{}^{\mu \nu
}=0,\qquad Q_H^{VV}{}^{\mu \nu }{}_\delta {}=-D_H{}_\delta {}g^{VV}{}^{\mu
\nu }=0
$$
(3) From Eq. \ref{hT-h.relation} comes the relation%
$$
g^{VV}{}^{\rho \nu }{}h_H^T{}_\rho {}^\mu {}_\delta {}-g^{HH}{}^{\mu \rho
}{}h_H{}^\nu {}_{\rho \delta }{}=0
$$
while the complement of Eq. \ref{hT-h.relation} yields the complementary
relation.

With a normal projection tensor field, indexes can be raised and lowered on
restricted tensors by using the intrinsic metrics so that the above relation
becomes just%
$$
h_H^T{}^{\nu \mu }{}_\delta =h_H{}^{\nu \mu }{}_\delta {}
$$
and its complement. Thus, some of the possible restricted metric tensors
have been set to zero by the normality conditions themselves (Eq. \ref
{normality}) half of the possible projection curvature tensors have been set
equal to the other half and we have complete freedom to raise and lower the
indexes of restricted tensors.

When $H$ is normal, the subspace $VT_P$ is completely determined once the
subspace $HT_P$ is given (provided that $HT_P$ is not null). Furthermore,
for a group with non-null orbits, the normal projection onto the orbits is
unique and obviously group-invariant. The remainder of this paper assumes
normal projection tensor fields unless stated otherwise.

\subsubsection{Projection curvatures}

The earlier results for arbitrary projection tensor fields may be
specialized to the case of normal projections onto group orbits. Eq. \ref
{h.TV.killing.eqn}, yields the condition $h_V^T{}_\sigma {}^{\left( \mu \nu
\right) }{}=0$ so that only the twist part of the $V$-projection curvature
remains.%
$$
h_V^T{}_\sigma {}^{\left[ \mu \nu \right] }{}=h_V{}_\sigma {}^{\left[ \mu
\nu \right] }{}=\omega _V{}_\sigma {}^{\mu \nu }{}
$$

Because the only non-zero projection tensors are $h_H{}$ and $\omega _V$ we
have the luxury of omitting the projection subscripts and defining%
$$
h{}^\nu {}_{\mu \delta }{}=h_H{}^\nu {}_{\mu \delta }{},\qquad \omega
{}^{\,\nu }{}_{\mu \delta }=\omega _V{}^\nu {}_{\mu \delta }{}.
$$
Since $H$ is surface-forming, there is no $H$-twist part of $h_H$. However
it still has the decomposition%
$$
h{}^\nu {}_{\mu \delta }{}=\sigma {}^\nu {}_{\mu \delta }{}+\frac 1s\theta
^\nu {}g_{HH}{}_{\mu \delta }
$$
where $\sigma {}^\nu {}_{\mu \delta }{}$ is the shear tensor defined by%
$$
\sigma {}^\nu {}_{\mu \delta }{}=h{}^\nu {}_{\mu \delta }{}-\frac 1s%
g_{HH}{}_{\mu \delta }h{}^{\nu \sigma }{}_\sigma {}
$$
and $\theta ^\nu {}=h{}^{\nu \sigma }{}_\sigma {}$ is the divergence.%
$$
{}.
$$

\subsubsection{Torsion and Ricci curvatures}

For this type of projection tensor field, the vanishing of the full torsion
tensor leads to the restricted torsion tensors
$$
S_{HH}^V{}^\rho {}_{\mu \nu }{}=0{},\qquad S_{HV}^H{}^\rho {}_{\mu \nu
}{}=h{}_\nu {}^\rho {}_\mu {}
$$
$$
S_{VV}^H{}^\rho {}_{\mu \nu }{}=2\omega {}^\rho {}_{\mu \nu }{},\qquad
S_{VH}^V{}^\rho {}_{\mu \nu }{}=\omega {}_\nu {}^\rho {}_\mu {}
$$
The Ricci curvature tensor decomposition then becomes
\begin{equation}
\label{Ricci.HH}R\left[ _{HH}\right] {}_{\alpha \beta }=R_{HH\,}^H{}_{\alpha
\beta }+D_{V\,\sigma }\,h^\sigma {}_{\alpha \beta }{}-\,h^\sigma {}_{\alpha
\beta }{}\theta {}_\sigma -\omega {}_\alpha {}^\rho \!_\sigma \,{}{}\omega
{}_\beta {}^\sigma {}_\rho ,
\end{equation}
$$
R\left[ _{VV}\right] {}_{\alpha \beta }=R_{VV\,}^V{}_{\alpha \beta
}+D_{V\,\beta }\,\theta {}_\alpha +D_{H\,\sigma }\,\omega {}^{\,\sigma
}{}_{\alpha \beta }{}-h{}_\alpha {}^\rho \!_\sigma \,h_\beta {}^\sigma
{}_\rho {}{},
$$
$$
R\left[ _{HV}\right] {}_{\alpha \beta }=R_{VH\,}^H{}_{\alpha \beta
}-D_{V\,\sigma }\,\omega {}_\alpha {}^\sigma \!_\beta {}-h{}^\rho {}_{\alpha
\sigma }{}\,\omega {}^\sigma \!_{\beta \rho }{}+\theta {}_\sigma {}\omega
{}_\alpha {}^\sigma \!_\beta {},
$$
$$
R\left[ _{VH}\right] {}_{\alpha \beta }=R_{HV\,}^V{}_{\alpha \beta
}-D_{H\,\sigma }\,h{}_\alpha {}^\sigma \!_\beta {}+D_{H\,\beta }\,\theta
{}_\alpha -\omega {}^{\,\rho }{}_{\alpha \sigma }{}h^\sigma {}_{\rho \beta
}{}.
$$

\subsubsection{Torsion Bianchi identities}

The cross-projected curvatures which appear in the last two equations are
unfamiliar and the expression for $R\left[ _{VV}\right] $ is not manifestly
symmetric in its indexes. The projected Torsion Bianchi identities solve
these problems. Specialize the identities given by Eq. \ref{T-Bianchi-asm}
to the case of normal projections onto group orbits and contract them where
possible to obtain the identities

$$
R_{HV}^{\!V}{}_{\nu \mu }=-D_{V\rho }\,{}\omega {}_\mu {}^\rho {}_\nu
{}+\omega {}_\sigma {}^\rho {}_\nu {}\,h{}_\rho {}^\sigma {}_\mu {},
$$
$$
R_{VH}^{\!H}{}_{\mu \gamma }{}=D_{H\mu }\,\theta {}_\gamma {}-D_{H\rho
}\,h{}_\gamma {}^\rho {}_\mu {}+h{}_\sigma {}^\rho {}_\mu {}\,\omega {}_\rho
{}^\sigma {}_\gamma {}-\theta {}_\sigma {}{}\,\omega {}_\mu {}^\sigma
{}_\gamma {},
$$
$$
D_{V[\gamma }\,{}\theta {}_{\nu ]}{}{}+D_{H\rho }\,\omega {}^{\,\rho
}{}_{\nu \gamma }{}=0.
$$
With the help of these identities, the cross-projections of the full Ricci
tensor become fully explicit,
\begin{equation}
\label{Ricci-hv-final}
\begin{array}{c}
R\left[ _{VH}\right] {}_{\alpha \beta }=R\left[ _{VH}\right] {}_{\alpha
\beta }= \\
-D_{V\rho }\,{}\omega {}_\beta {}^\rho {}_\alpha {}+D_{H\,\beta }\,\theta
{}_\alpha {}-D_{H\,\sigma }\,h{}_\alpha {}^\sigma \!_\beta {}+2\omega
{}_\sigma {}^\rho {}_\alpha {}\,h{}_\rho {}^\sigma {}_\beta ,
\end{array}
\end{equation}
while the normal projection becomes manifestly symmetric
\begin{equation}
\label{Ricci-VV-final}R\left[ _{VV}\right] {}_{\alpha \beta
}=R_{VV\,}^V{}_{\alpha \beta }+D_{V\,(\beta }\,\theta {}_{\alpha
)}-h{}_\alpha {}^\rho \!_\sigma \,h_\beta {}^\sigma {}_\rho {}.
\end{equation}

The remaining nontrivial projected torsion Bianchi identity cannot be
contracted but takes the form of an equation of motion for the $V$-twist
tensor.
\begin{equation}
\label{twist.bianchi}D_{V[\gamma }\,{}\omega {}^{\,\rho }{}_{\mu \nu
]}{}+\omega {}^{\,\sigma }{}_{[\mu \nu }{}h{}_{\gamma ]}{}^\rho {}_\sigma
\,=0
\end{equation}
In Kaluza-Klein theories, the $V$-twist becomes the electromagnetic field
tensor and this identity becomes the source-free Maxwell's equation\cite
{Kaluza-Klein-gen}.

\subsubsection{Normal orbit projections of Einstein's equations}

Equations \ref{Ricci.HH}, \ref{Ricci-hv-final} and \ref{Ricci-VV-final} may
be used to obtain Einstein's equations. It is convenient to decompose the
resulting equations into trace and trace-free parts and to re-organize the
trace parts so that Einstein's equations become
\begin{equation}
\label{Einst-Rvv}\left( 2-s\right) R_{HH\,}^H{}-sR_{VV\,}^V+2\left(
1-s\right) D\cdot \theta {}-\left( 1-s\right) \theta ^2+\left( 2-s\right)
\omega ^{\,2}+s\sigma ^2=16\pi \kappa T{}_{HH}{}
\end{equation}
\begin{equation}
\label{Einst-div}R_{HH\,}^H+D\cdot \theta {}-\theta ^2+\omega ^{\,2}=\frac{%
8\pi \kappa }{2-d}\left( sT{}_{VV}{}+\left( 2-d+s\right) T{}_{HH}{}\right)
\end{equation}
\begin{equation}
\label{Einst-TF-HH}%
\mathop{\rm TF}
R_{HH\,}^H{}_{\alpha \beta }+D_{V\,\sigma }{}\sigma {}^\sigma {}_{\alpha
\beta }{}-\theta {}_\sigma {}\sigma {}^\sigma {}_{\alpha \beta }{}-%
\mathop{\rm TF}
\omega {}_\alpha {}^\rho \!_\sigma \,{}{}\omega _\beta {}^\sigma {}_\rho
=8\pi \kappa
\mathop{\rm TF}
T{}_{HH}{}_{\alpha \beta }
\end{equation}
\begin{equation}
\label{Einst-TF-VV}%
\mathop{\rm TF}
R_{VV\,}^V{}_{\alpha \beta }+%
\mathop{\rm TF}
D_{V\,(\beta }\,\theta {}_{\alpha )}{}-%
\mathop{\rm TF}
\sigma {}_\alpha {}^\rho \!_\sigma \,{}\sigma {}_\beta {}^\sigma {}_\rho {}-%
\frac 1s%
\mathop{\rm TF}
\theta {}_\alpha {}{}\theta {}_\beta {}=8\pi \kappa
\mathop{\rm TF}
T{}_{VV}{}_{\alpha \beta }
\end{equation}
\begin{equation}
\label{Einst-HV}
\begin{array}{c}
-D_{V\rho }\,{}\omega {}_\beta {}^\rho {}_\alpha {}+\left( 1-
\frac 1s\right) D_{H\,\beta }\,\theta {}_\alpha -D_{H\,\sigma }\,\sigma
{}_\alpha {}^\sigma \!_\beta {} \\ +2\omega {}_\sigma {}^\rho {}_\alpha
{}\,\sigma {}_\rho {}^\sigma {}_\beta +\frac 2s\omega {}_\beta {}^\rho
{}_\alpha {}\,\theta {}_\rho {}=8\pi \kappa T{}_{VH}{}_{\alpha \beta }
\end{array}
\end{equation}
with the abbreviations%
$$
\theta ^2=\theta {}_\sigma \theta ^\sigma ,\qquad \omega ^{\,2}=-\omega
{}_\tau {}^\rho \!_\sigma \,{}\omega {}^{\tau \sigma }{}_\rho {},\qquad
\sigma ^2=\sigma {}_\tau {}^\rho \!_\sigma \,{}\sigma {}^{\tau \sigma
}{}_\rho {},\qquad D\cdot \theta =D_{V\,\sigma }\theta ^\sigma
$$
and the notation TF for the trace-free part of a second-rank tensor.

Equation \ref{Einst-div} is the result of combining the trace equations and
has two remarkable properties: (1) It does not involve the shear. (2) All of
the dependence on dimensionality is associated with the stress-energy
components. Because this one Einstein equation is the same for any spacetime
with an isometry, it is worth examining more closely. In terms of the
orbit-area $a$, the remarkable Eq. \ref{Einst-div} becomes even more
remarkable:
\begin{equation}
\label{orbit-area-eqn}-\Delta _Va+R_{HH\,}^Ha=\frac{8\pi \kappa }{2-d}\left(
sT{}_{VV}{}+\left( 2-d+s\right) T{}_{HH}{}\right) a
\end{equation}
where
$$
\Delta _Vf=g^{VV}{}^{\alpha \beta }D_{V\alpha }D_{V\beta }f
$$
defines the harmonic operator $\Delta _V$ on the quotient space. In this
form, I will refer to this combination of Einstein's equations as the {\em %
orbit-area equation}. This equation determines the behavior of the
orbit-area function which, in turn, determines the global topology of
spacetime.\cite{2-param-1}

\subsection{Isotropic orbits}

\subsubsection{Normality constraints}

{}From the projection $g_{HV}$ and a group-invariant vector field $v$
construct the group-invariant form $g_{HV}{}_{\sigma \rho }{}v{}^\rho {}$
and then construct a forbidden orbit-tangent vector field by using $g^{HH}$.
The resulting constraint%
$$
g^{HH}{}^{\alpha \sigma }g_{HV}{}_{\sigma \beta }=0
$$
implies that the projection is normal whenever the group-orbits are non-null
(so that $g^{HH}$ can have an inverse). Similarly, construct the
group-invariant form $g_{VV}{}_{\sigma \rho }{}v^\rho $ and then a forbidden
orbit-tangent vector field $g^{HV}{}^{\alpha \sigma }{}g_{VV}{}_{\sigma \rho
}{}v^\rho $ to find the constraint%
$$
g^{HV}{}^{\alpha \sigma }{}g_{VV}{}_{\sigma \rho }{}=0
$$
which implies that the projection is normal whenever the quotient geometry
is non-null.

\subsubsection{Projection-curvature constraints}

The simplest invariant vector field to construct is $h_V{}^\alpha {}_{\rho
\sigma }v^\rho u^\sigma $ where $u,v$ are invariant vector fields. Because
the resulting vector field lies in $HT_P$ and is tangent to the group-orbit,
it can only be zero. The invariant group evolution assumption means that the
vector fields $u,v$ can be chosen arbitrarily from a set which spans $VT_P$
at any one point $P$ so the projection curvature obeys the constraint $%
h_V{}^\alpha {}_{\rho \sigma }=0$. One consequence of this constraint is $%
\omega _V{}^\alpha {}_{\rho \sigma }=0$ which implies that $V$ is not only a
normal projection but is surface-forming. Thus, isotropic orbits are always
surface-orthogonal.

The projection curvature $h_H^T$ requires a less direct approach. From $h_H^T$
one can construct a symmetric tensor field
$$
t\left( v\right) {}_{\alpha \beta }{}=v^\tau h_H^T{}_\tau {}^\rho
{}_{(\alpha }{}g_{HH}{}_{\beta )\rho }
$$
and then seek solutions to the eigen-value equation $t\left( v\right)
{}_{\alpha \beta }{}u^\beta =\lambda g_{HH}{}_{\alpha \rho }{}u^\rho $. If
there are distinct eigenvalues, then there will be distinct eigenvectors
which constitute group-invariant vector-fields which are tangent to the
group orbits. To avoid this forbidden possibility, require $t\left( v\right)
{}_{\alpha \beta }{}=\alpha \left( v\right) g_{HH}{}_{\alpha \beta }{}$ or%
$$
h_H^T{}_\tau {}^\rho {}_{(\alpha }{}g_{HH}{}_{\beta )\rho }=\alpha _\tau
g_{HH}{}_{\alpha \beta }{}
$$
where $\alpha \in V^{*}\hat T_P$. Similarly, the projection curvature $h_H$
can be used to construct a second-rank group-invariant tensor field whose
eigenvectors are the forbidden orbit-tangent invariant vector fields. The
resulting constraint is%
$$
g_{VV}{}_{\tau \rho }{}h_H{}^\rho {}_{\left( \alpha \beta \right) }=\beta
_\tau g_{HH}{}_{\alpha \beta }
$$
where $\beta \in V^{*}\hat T_P.$

\subsubsection{Summary of consequences of isotropic orbits}

Let $V_d$ be a geometry which admits an isometry group $G_n$ and take $H$ to
be a group-invariant projection tensor field which projects vectors tangent
to $s$-dimensional orbits in the family $\left[ V_d/G_n\right] \left(
s\right) $. From the previous section, this projection is necessarily normal
and surface-orthogonal which means that cross-projected metric terms such as
$g_{VH}{}_{\alpha \beta }{}$ are all zero, the projected metrics $%
g_{HH},g^{HH},g_{VV},g^{VV}$ may be used to raise and lower indexes (of the
appropriate projection type) on tensors, and the projection curvatures $h$
and $h^T$ are the same except for index placement (which is now easy to
change). The possible values of the projection curvatures are severely
constrained and must have the form%
$$
\begin{array}{c}
h_V^T{}^\alpha {}_{\rho \sigma }=h_V{}^\alpha {}_{\rho \sigma }=0 \\
h_H^T{}^\rho {}_{\alpha \beta }=h_H{}^\rho {}_{\alpha \beta }=h{}^\rho
{}_{\alpha \beta }=\alpha ^\rho g_{HH}{}_{\alpha \beta }
\end{array}
$$
where $\alpha $ is a vector in $VT_P$. From the definition of the
divergence, $\theta _\rho $ it is easy to see that $\theta _\rho =s\alpha
_\rho $ or
\begin{equation}
\label{h-iso}h{}^\rho {}_{\alpha \beta }=\frac 1s\theta ^\rho
g_{HH}{}_{\alpha \beta }
\end{equation}
and
\begin{equation}
\label{shear-free}\sigma {}^\rho {}_{\alpha \beta }=0.
\end{equation}

\subsubsection{Effective orbit-size parameter}

Each orbit can be characterized by an effective size parameter $r$ which is
defined by requiring the orbit area to be given by $a=Br^s$ for some
constant $B$. For example, if the group is SO$\left( 2\right) $ then $s=1$,
the orbits are circles, $a$ can be chosen to be the total circumference of
an orbit, and the choice $B=2\pi $ makes $r$ the radius which an orbit would
have in a flat embedding space. Similarly, for SO$\left( 3\right) $ the
orbits are two-spheres and the choice $B=4\pi ^2$ makes $r$ again the flat
embedding radius of an orbit --- sometimes called the luminosity radius.
When $s=3$, the value of $r$ becomes a measure of the size of the universe
in a cosmological model. Because $\theta {}_\alpha {}$ depends only on the
logarithm of $a$, the choice of the constant $B$ is of no immediate
consequence and the divergence takes the form
\begin{equation}
\label{div-orbit-size}\theta {}_\alpha {}=-a^{-1}D_{V\alpha
}a=-sr^{-1}D_{V\alpha }r.
\end{equation}

The scalar curvature $R_{HH}^H{}$ of an orbit may also be expressed in terms
of the effective size parameter $r$. When the Riemann curvature tensor of an
orbit is expressed in a basis which is constructed from the group
coordinates, its components are determined solely by the underlying group
and are the same for all orbits. Since the Ricci tensor of an orbit is
simply a contraction of the Riemann tensor without any use of the metric
tensor, it too has group-basis components which are the same for all orbits.
The scalar curvature $R_{HH}^H{}=g^{HH}{}^{\alpha \beta }R_{HH}^H{}_{\alpha
\beta }$ therefore depends on the orbit only through the inverse metric
components $g^{HH}{}^{\alpha \beta }{}=\omega ^{H\alpha }\bullet \omega
^{H\beta }$ where the forms $\omega ^{H\alpha }$ are defined in terms of the
underlying group parameters (typically angles). Thus, one always has
\begin{equation}
\label{R-HHH.orbit-size}R_{HH}^H{}=\frac{K_{\text{G}}}{r^2}.
\end{equation}
The constant $K_{\text{G}}$ is determined by the group and by the choices
which define the effective size function $r$. A simple way to calculate $K_{%
\text{G}}$ is to consider the orbits which are generated when the group acts
on a flat spacetime manifold. For the groups SO$\left( 2\right) $, SO$\left(
3\right) $, and SO$\left( 4\right) $ the constants are found to be $K_{\text{%
G}}=0,2,6$.

\section{Applications}
\label{apps}

\subsection{Einstein's equations for isotropic spacetimes}

With the simplifications which are afforded by Eqs. \ref{shear-free}, \ref
{div-orbit-size}, and \ref{R-HHH.orbit-size}, two of the trace-decomposed,
normal orbit projections of Einstein's equations (Eqs. \ref{Einst-TF-HH} and
\ref{Einst-HV}) become constraints on the stress-energy tensor while the
remaining equations (Eqs. \ref{Einst-Rvv}, \ref{Einst-div}, and \ref
{Einst-TF-VV}), become
\begin{equation}
\label{RV-r}
\begin{array}{c}
s\left( 1-s\right) \left[ \left( 2-s\right) r^{-2}\left( D_{V\sigma
}r\right) \left( D_V{}^\sigma r\right) -2r^{-1}D_{V\,\sigma }\,D_V{}^\sigma
r\right] \\
+\left( 2-s\right) r^{-2}K_{\text{G}}=sR_{VV\,}^V+16\pi \kappa T_{HH}{}
\end{array}
\end{equation}
\begin{equation}
\label{scalar-r}
\begin{array}{c}
s\left[ \left( 1-s\right) r^{-2}\left( D_{V\sigma }r\right) \left(
D_V{}^\sigma r\right) -r^{-1}D_{V\,\sigma }\,D_V{}^\sigma r\right] +
\frac{K_{\text{G}}}{r^2} \\ =8\pi \kappa \left[ \frac{d-s-2}{d-2}T_{HH}{}-%
\frac s{d-2}T_{VV}{}\right]
\end{array}
\end{equation}
\begin{equation}
\label{TF-r}
\begin{array}{c}
-s
\mathop{\rm TF}
\left[ \,r^{-1}D_{V\,\beta }D_{V\alpha }r-2r^{-2}\left( D_{V\,\beta
}\,r\right) \left( D_{V\alpha }r\right) \right] \\ =8\pi \kappa
\mathop{\rm TF}
T_{VV}{}_{\alpha \beta }{}-%
\mathop{\rm TF}
R_{VV\,}^V{}_{\alpha \beta }
\end{array}
\end{equation}
These three equations form a complete system for determining the orbit size
function $r$ and the quotient geometry. Notice that Eq. \ref{TF-r} does not
depend on the metric tensor in any way except through the connection. It's
sole function is to determine the quotient space connection $D_V$. Equations
\ref{scalar-r} and \ref{RV-r} determine the metric and matter variables
including the orbit size function. However, as will be seen in these
examples, the exact way that this system of equations functions is strongly
affected by the ''accidents'' of dimension.

\subsubsection{Codimension = 1: Friedman-Robertson-Walker cosmologies}

For $d-s=1$ the trace-free equation (Eq. \ref{TF-r}) is identically
satisfied and the quotient-space connection is described by identifying a
proper time function $\tau $ and taking the single orthonormal-frame
components of the restricted derivative to be just $D_{V0}f=df/d\tau $. The
only remaining equations to be solved are Eq. \ref{scalar-r} and Eq. \ref
{RV-r} which become the familiar cosmological equations for the radius $r$
of the universe.

As a specific and familiar example, consider the parameter values $d=4,\;s=3$
which correspond to isotropic, homogeneous cosmological models and take the
stress-energy tensor to be that of a perfect fluid.\cite{FRW-cosmology} One
''accident'' of this choice of dimensions is the simplicity of the perfect
fluid stress-energy tensor\cite{prjctn1}:
\begin{equation}
\label{stress-tensor-fluid}T{}^\mu {}_\nu {}=p_H{}H{}^\mu {}_\nu
{}+p_V{}V{}^\mu {}_\nu {}=pH{}^\mu {}_\nu {}-\rho {}V{}^\mu {}_\nu {}.
\end{equation}
To obtain a definite example which will be needed later, take the equation
of state to be that of incoherent radiation, $p=\rho /3$, so that the
projected traces which enter into Eqs. \ref{scalar-r} and \ref{RV-r} are%
$$
T_{HH}{}=\rho ,\qquad T_{VV}{}=-\rho .
$$
Choose a timelike unit vector $e_0=\partial /\partial \tau $ and denote
proper-time derivatives by dots so that Equations \ref{scalar-r} and Eq. \ref
{RV-r} become%
$$
3\frac{\ddot r}r+6\frac{\dot r^2}{r^2}+\frac{K_{\text{G}}}{r^2}=8\pi \kappa
\rho
$$
$$
-6\frac{\ddot r}r-3\frac{\dot r^2}{r^2}-\frac{K_{\text{G}}}{2r^2}=8\pi
\kappa \rho .
$$
These equations can be used in one of two ways. Subtracting them gives the
familiar second order equation for the radius of a radiation-dominated
universe\cite{FRW-rad-dom}
\begin{equation}
\label{rad-cosmology}\frac{\ddot r}r+\frac{\dot r^2}{r^2}+\frac{K_{\text{G}}%
}{6r^2}=0
\end{equation}
while eliminating the second derivative between them gives the initial value
constraint
\begin{equation}
\label{rad-iv-eqn}\frac{\dot r^2}{r^2}+\frac{K_{\text{G}}}{6r^2}=\frac 83\pi
\kappa \rho .
\end{equation}
The constant $K_{\text{G}}/6$ is the usual topology parameter (often denoted
$k$) which has values -1, 0, +1 for universes which are open, spatially
flat, and closed respectively.

\subsubsection{Codimension = 2: Spherical symmetry and Birkhoff's theorem}

For $d-s=2$ the Ricci curvature of the quotient space is proportional to the
metric tensor. The trace-free part of the Ricci curvature then vanishes. Set
the stress-energy tensor equal to zero in order to consider vacuum
spacetimes and focus on spacelike group orbits. For $s=2$, this case
includes the exterior metrics of spherically symmetric systems. The
trace-free part of Einstein's equations then collapses to just
\begin{equation}
\label{2dim-TF}%
\mathop{\rm TF}
D_{V\,\beta }D_{V\alpha }r=0
\end{equation}
This equation is supposed to determine the quotient geometry (or
equivalently the geometry of the reference surface which is perpendicular to
the group orbits) and it will be seen that it does that job admirably and
simply.

Introduce orthonormal basis vectors $\left( e_0,e_1\right) $ on $VT_P$,
aligned with the function $r$. Remember that these vector fields are
directional derivatives and use the notation $\left( e_1f\right) =f^{\prime
} $ where $f$ is any function on the reference surface. Focus on a spacetime
region where this function has a spacelike gradient so that, at each point
the basis vectors can be chosen so that $\left( e_0r\right) =0$ and $\left(
e_1r\right) =r^{\prime }>0$. Within some local region, the function $r$
provides one coordinate on the two-dimensional reference surface and there
will be another coordinate $t$ whose level-curves are the integral curves of
the perpendicular vector field $e_1$. In terms of these coordinates, the
orthonormal basis vectors are%
$$
e_0=N\frac \partial {\partial t},\qquad e_1=r^{\prime }\frac \partial {%
\partial r}.
$$
In general the functions $r^{\prime }$ and $N$ can depend on both
coordinates.

There are only two non-zero independent orthonormal-frame connection
coefficients:%
$$
\gamma _A=\Gamma ^V{}_{VV}{}^0{}_{1A}=\Gamma ^V{}_{VV}{}^1{}_{0A},\qquad
A=0,1.
$$
{}From the off-diagonal components of Eq. \ref{2dim-TF} come the two
conditions%
$$
\gamma _1r^{\prime }=0,\qquad N\frac{\partial r^{\prime }}{\partial t}=0
$$
which eliminate one connection component and ensure that $r^{\prime }$ is
independent of $t$ while the anti-trace combination of diagonal components
of Eq. \ref{2dim-TF} gives $-\gamma _0r^{\prime }+r^{\prime \prime }=0$
which can be solved for the other connection component.

Because the group-orbit projection tensor field is normal, the projected
torsion vector $S^V{}_{VV}{}^A{}_{01}e_A$ must vanish or%
$$
\left[ e_0,e_1\right] -\gamma _0e_0+\gamma _1e_1=0.
$$
This condition reduces to just one equation%
$$
-N^{\prime }-\frac{r^{\prime \prime }}{r^{\prime }}N=0
$$
which integrates to
$$
N\left( r,t\right) =\frac{N_0\left( t\right) }{r^{\prime }}.
$$
The remaining function $N_0$ can be absorbed by redefining $t$ so that $%
N\left( r,t\right) =1/r^{\prime }$.

To determine the rest of the geometry, specialize the orbit-area equation
(Eq. \ref{scalar-r}) to this case and obtain%
$$
-4\frac{r^{\prime \prime }}r-2\frac{r^{\prime 2}}{r^2}+\frac 2{r^2}=0
$$
which has the first integral $2rr^{\prime 2}-2r=-4m$ where the integration
constant $-4m$ has been chosen with a certain amount of forethought. Solve
this expression for $r^{\prime }$ and find the expressions for the
orthonormal basis vectors
$$
e_0=\frac 1{\sqrt{1-\frac{2m}r}}\frac \partial {\partial t},\qquad e_1=\sqrt{%
1-\frac{2m}r}\frac \partial {\partial r}.
$$

{}From the assumption that a vacuum spacetime has a group of motions with two
dimensional isotropic orbits and the added restriction to a region where the
orbit area function has a spacelike gradient, we have constructed the
Schwarzschild solution. Along the way an additional isometry group has
spontaneously appeared because this solution is static. This result is
usually called Birkhoff's Theorem.\cite{Birkhoff-Th}

\subsubsection{Codimension = 2: Five dimensional cosmology}

Consider a five-dimensional manifold $V_5$ with a metric of signature $%
\left( -++++\right) $ and a group of motions $G$ whose orbits are
three-dimensional spacelike surfaces. In this case, the parameters
introduced earlier are $s=3,\;d=5$. The quotient space $\left[ V_5/G\right]
\left( 3\right) $ is two dimensional so the analysis which worked so well
for spherically symmetric four-manifolds should work here as well. The group
orbits are now three dimensional homogeneous and isotropic spaces. A family
of these is a cosmology. Thus, we are looking for those solutions of
Einstein's vacuum equations in five dimensions which can describe
cosmological models. This time, assume that the orbit size function $r$ is
timelike.

In $VT_P$ (or equivalently, in the quotient space $V_5/G$) introduce the
orthonormal basis $\left\{ e_A\right\} $ with
$$
\begin{array}{c}
e_0\cdot e_0=-1,\qquad e_1\cdot e_1=1 \\
e_0r=e_0\left( r\right) =\dot r,\qquad e_1r=e_1\left( r\right) =0
\end{array}
$$
and%
$$
e_0=\dot r\frac \partial {\partial r},\qquad e_1=N\frac \partial {\partial
\chi }
$$
Notice that the unit vector $e_0$ can also be written in terms of the proper
time $\tau $ as $e_0=\partial /\partial \tau $. Thus, dots denote
proper-time derivatives.

The geometry of the quotient space is determined, as before, by Eq. \ref
{2dim-TF} which now consists of the conditions

$$
N\frac{\partial \dot r}{\partial \chi }=0,\qquad \gamma _0=0,\qquad \ddot r%
-\gamma _1\dot r=0
$$
The condition that the projected torsion vanishes leads, as before, to the
expression $N=1/\dot r$. All of these quantities are independent of the
fifth dimensional coordinate $\chi $. We have a surprise symmetry just as
happened in the 4D case. It remains only to use the orbit-size equation (Eq.
\ref{scalar-r}) to determine the behavior of $r$. In this case, the equation
takes a familiar form and is just Eq. \ref{rad-cosmology} --- the equation
which governs a radiation-dominated universe. Working backward to the
effective 4D stress-energy tensor can only yield the stress-energy of
incoherent radiation --- a result which has been noticed before in
spherically symmetric solutions to the 5D Einstein equations.\cite
{5D-spher-sols,kklein-cosm-sol}

In this curious parallel to Birkhoff's theorem, the requirement that a five
dimensional vacuum spacetime obey the Cosmological Principal by having
three-dimensional space sections which are homogeneous and isotropic gives
rise to a spontaneous additional symmetry which suppresses a dimension.\cite
{Birkhoff-multidim} The resulting solution can be regarded as a radiation
dominated four-dimensional cosmological model with the radiation pressure
supplied by a scalar field --- a manifestation of the suppressed fifth
dimension.\cite{kklein-cosm-sol} Distances in that suppressed direction
scale as the function $N^{-1}$ which is proportional to the expansion rate $%
\dot r$. Thus, as the universe evolves and the expansion rate slows, the
fifth dimensional size of the universe collapses. This picture fits the
usual scenario of a spontaneous dimensional reduction.\cite{dim-reduction}
The curious (and, as far as I can tell, new) aspect of this picture is that
the dimensional reduction is not an independent contrivance of the initial
conditions but a necessary accompaniment of those initial conditions which
produce homogeneity and isotropy.

\section{Discussion}

The projection tensor framework which has been developed here offers two
main advantages: (1) It is metric-independent and works for any connection.
(2) It works for any combination of dimensions. The examples which have been
developed in this paper have mostly exploited the second advantage,
analyzing systems with arbitrary dimensionalities and then considering the
accidents which happen when certain combinations of dimensions are chosen.
The first advantage, metric-independence, has been exploited only in part.
It is clearly useful to have metric-independent geometrical structures such
as the divergence and twist of a projection-tensor field. The examples show
how these structures can play such central roles in the Einstein equations
that the spacetime metric tensor fades into the background. However, all of
the examples rely upon normal projection tensor fields.

The normality condition, which depends directly upon the metric, is
extremely powerful. It cuts the number of independent projection curvatures
in half and makes the restricted covariant derivatives metric compatible on
their respective subspaces and even between subspaces. The tilted
projection-tensor formalism which has been developed here makes it possible
to do without the normality condition and thus makes it possible to search
for useful alternative conditions. A situation where an alternative
condition would clearly be useful is projections onto null surfaces.\cite
{dual-null-formalism} Another situation which calls for alternative
conditions is the analysis of ''non-projective'' Kaluza-Klein unified field
theories which are most naturally stated in terms of tilted projection
tensor fields.

The applications have been restricted to isotropic spacetimes in order to
provide simple and familiar examples. It is quite easy to extend the
applications to include groups with anisotropic orbits and, in this way,
revisit Kasner universes, Mixmaster universes, cylindrical waves, and the
rotating star problem. The basic pattern for solving Einstein's equations
remains the same in all of these cases: (1) Solve the orbit-area equation
(Eq.~\ref{orbit-area-eqn}) for the orbit-area function, thus fixing the
over-all topology of the spacetime and choosing a coordinate. (2) Solve the
orbit-tangent, trace-free equation (Eq.~\ref{Einst-TF-HH}) (a wave equation
on the quotient space) for the shear which carries the gravitational wave
degrees of freedom. (3) Solve the orbit-orthogonal, trace-free equation (Eq.~%
\ref{Einst-TF-VV}) for the restricted connection on the space of orbits and
in this way determine the geometry of the quotient space. In the solvable
cases, these steps decouple cleanly from one another much as they do in the
simple isotropic examples.

\end{document}